\documentclass[twocolumn, twocolappendix]{aastex631}
\usepackage[T5]{fontenc}
\usepackage[utf8]{inputenc}
\usepackage{svg}
\usepackage{gensymb}
\usepackage{amsmath}
\usepackage{rotating}
\def\msun{\,{\rm M_\odot}}

\usepackage[dvipsnames]{xcolor}
\definecolor{mycyan}{rgb}{0.5, 1.0, 1.0}
\definecolor{myorange}{rgb}{1.0, 0.824, 0.5}
\definecolor{myred}{rgb}{1.0, 0.5, 0.5}
\definecolor{mydarkred}{rgb}{0.773, 0.5, 0.5}

\defcitealias{Kim+2013}{I}
\defcitealias{Kim+2016}{II}
\defcitealias{Roca-Fabrega+2021}{III}
\defcitealias{Roca-Fabrega+2024}{IV}
\defcitealias{Jung+2024}{V}
\defcitealias{Strawn+2024}{VI}
\defcitealias{Rodriguez-Cardoso+2025a}{VII}
\defcitealias{Jung+2025a}{VIII}
\defcitealias{Kim+2025}{X}
\defcitealias{Barrow+2026a}{XI}

\begin{document}

\title{The \textit{AGORA} High-resolution Galaxy Simulations Comparison Project. IX - Part 1. Effects of a Major Galaxy Merger on Star Formation of a Milky Way-mass Galaxy Progenitor}
\author[0009-0002-2290-8039]{Thịnh Hữu Nguyễn}
\altaffiliation{Corresponding author}
\affiliation{Department of Astronomy, University of Illinois at Urbana-
Champaign, Urbana, IL 61801, USA; thinhhn2@illinois.edu}
\affiliation{Center for AstroPhysical Surveys, National Center for Supercomputing Applications, Urbana, IL 61801, USA}

\author[0000-0002-8638-1697]{Kirk S. S. Barrow}
\affiliation{Department of Astronomy, University of Illinois at Urbana-
Champaign, Urbana, IL 61801, USA}

\author[0000-0002-9144-1383]{Minyong Jung}
\altaffiliation{Corresponding author}
\affiliation{Center for Theoretical Physics, Department of Physics and Astronomy, Seoul National University, Seoul 08826, Republic of Korea; wispedia@snu.ac.kr}

\author[0000-0002-9158-195X]{Ram\'{o}n Rodr\'{i}guez-Cardoso}
\altaffiliation{Corresponding author}
\affil{Departamento de Física de la Tierra y Astrofísica, Fac. de C.C. Físicas, Universidad Complutense de Madrid, E-28040 Madrid, Spain; ramorodr@ucm.es}
\affil{Instituto de Física de Partículas y del Cosmos, IPARCOS, Fac. C.C. Físicas, Universidad Complutense de Madrid, E-28040 Madrid, Spain}

\author[0000-0002-6299-152X]{Santi Roca-F\`{a}brega}
\altaffiliation{Code leaders}
\affil{Lund Observatory, Division of Astrophysics, Department of Physics, Lund University, SE-221 00 Lund, Sweden}
\affil{Departamento de F\'{i}sica de la Tierra y Astrof\'{i}sica, Facultad de Ciencias F\'{i}sicas, Plaza Ciencias, 1, 28040 Madrid, Spain}

\author[0000-0003-4464-1160]{Ji-hoon Kim}
\altaffiliation{Code leaders}
\affiliation{Seoul National University Astronomy Research Center, Seoul 08826, Korea}
\affiliation{Center for Theoretical Physics, Department of Physics and Astronomy, Seoul National University, Seoul 08826, Korea}
\affiliation{Institute for Data Innovation in Science, Seoul National University, Seoul 08826, Republic of Korea}

\author[0000-0001-5091-5098]{Joel R. Primack}
\altaffiliation{Deceased}
\affil{Department of Physics, University of California at Santa Cruz, Santa Cruz, CA 95064, USA}

\author[0000-0001-7457-8487]{Kentaro Nagamine}
\altaffiliation{Code leaders}
\affiliation{Theoretical Astrophysics, Department of Earth and Space Science, Graduate School of Science, Osaka University, Toyonaka, Osaka, 560-0043, Japan}
\affiliation{Theoretical Joint Research, Forefront Research Center, Graduate School of Science, The University of Osaka, 1-1 Machikaneyama, Toyonaka, Osaka 560-0043, Japan}
\affiliation{Kavli IPMU (WPI), UTIAS, The University of Tokyo, Kashiwa, Chiba 277-8583, Japan}
\affiliation{Department of Physics \& Astronomy, University of Nevada Las Vegas, Las Vegas, NV 89154, USA}
\affiliation{Nevada Center for Astrophysics, University of Nevada, Las Vegas, 4505 S. Maryland Pkwy, Las Vegas, NV 89154-4002, USA}

\author[0000-0001-8531-9536]{Renyue Cen}
\affil{Center for Cosmology and Computational Astrophysics, Institute for Advanced Study in Physics, Zhejiang University, Hangzhou 310027, People's Republic of China}
\affil{Institute of Astronomy, School of Physics, Zhejiang University, Hangzhou 310027, People's Republic of China}

\author[0000-0002-8680-248X]{Daniel Ceverino}
\affil{Universidad Aut\'{o}noma de Madrid, Ciudad Universitaria de Cantoblanco, E-28049 Madrid, Spain}
\affil{CIAFF, Facultad de Ciencias, Universidad Aut\'{o}noma de Madrid, E-28049 Madrid, Spain}

\author[0000-0002-2113-4863]{Weiguang Cui}
\affil{Institute for Astronomy, Royal Observatory, Edinburgh EH9 3HJ, UK}
\affil{Departamento de Física Teórica, Universidad Autónoma de Madrid, Módulo 15, E-28049 Madrid, Spain}
\affil{Centro de Investigación Avanzada en Física Fundamental (CIAFF), Facultad de Ciencias, Universidad Autónoma de Madrid, E-28049 Madrid, Spain}

\author[0000-0003-0073-3012]{Anna Genina}
\altaffiliation{Code leaders}
\affil{Institute for Astronomy, University of Edinburgh, Royal Observatory, Blackford Hill, Edinburgh EH9 3HJ, UK}

\author[0000-0002-7820-2281]{Hyeonyong Kim}
\altaffiliation{Code leaders}
\affiliation{Center for Theoretical Physics, Department of Physics and Astronomy, Seoul National University, Seoul 08826, Korea}

\author[0000-0002-5712-6865]{Yuri Oku}
\affiliation{Theoretical Astrophysics, Department of Earth and Space Science, Graduate School of Science, Osaka University, Toyonaka, Osaka, 560-0043, Japan}
\affil{Center for Cosmology and Computational Astrophysics, Institute for Advanced Study in Physics, Zhejiang University, Hangzhou 310027, People's Republic of China}

\author[0000-0002-3764-2395]{Johnny W. Powell}
\altaffiliation{Code leaders}
\affil{Department of Physics, Reed College, Portland, OR 97202, USA}

\author[0000-0002-6227-0108]{Yves Revaz}
\altaffiliation{Code leaders}
\affil{Institute of Physics, Laboratoire d'Astrophysique, \'{E}cole Polytechnique F\'{e}d\'{e}rale de Lausanne (EPFL), CH-1015 Lausanne, Switzerland}

\author[0009-0002-1398-6537]{Pablo Granizo}
\affiliation{Theoretical Astrophysics, Department of Earth and Space Science, Graduate School of Science, Osaka University, Toyonaka, Osaka, 560-0043, Japan}
\affil{Universidad Aut\'{o}noma de Madrid, Ciudad Universitaria de Cantoblanco, E-28049 Madrid, Spain}

\author[0000-0001-6106-7821]{Alessandro Lupi}
\altaffiliation{Code leaders}
\affil{Como Lake Center for Astrophysics, DiSAT, Universit\`a degli Studi dell'Insubria, via Valleggio 11, I-22100 Como, Italy}
\affil{INFN, Sezione di Milano-Bicocca, Piazza della Scienza 3, I-20126 Milano, Italy}
\affil{INAF - Osservatorio di Astrofisica e Scienza dello Spazio di Bologna, Via Gobetti 93/3, I-40129 Bologna}

\author{Ikkoh Shimizu}
\altaffiliation{Code leaders}
\affil{Shikoku Gakuin University, 3-2-1 Bunkyocho, Zentsuji, Kagawa, 765-8505, Japan}

\author{H\'{e}ctor Vel\'{a}zquez}
\altaffiliation{Code leaders}
\affil{Instituto de Astronom\'{i}a, Universidad Nacional Aut\'{o}noma de M\'{e}xico, A.P. 70-264, 04510, Mexico, D.F., Mexico}

\author[0000-0002-5969-1251]{Tom Abel}
\affil{Kavli Institute for Particle Astrophysics and Cosmology, Stanford University, Stanford, CA 94305, USA}
\affil{Department of Physics, Stanford University, Stanford, CA 94305, USA}
\affil{SLAC National Accelerator Laboratory, Menlo Park, CA 94025, USA}

\author[0000-0002-4287-1088]{Oscar Agertz}
\affil{Lund Observatory, Division of Astrophysics, Department of Physics, Lund University, SE-221 00 Lund, Sweden}

\author[0000-0003-4174-0374]{Avishai Dekel}
\altaffiliation{Deceased}
\affil{Center for Astrophysics and Planetary Science, Racah Institute of Physics, The Hebrew University, Jerusalem 91904, Israel}

\author[0000-0003-4597-6739]{Boon Kiat Oh}
\affiliation{Department of Physics, University of Connecticut, U-3046, Storrs, CT 06269, USA}
\affiliation{School of Physics, Korea Institute for Advanced Study, 85 Hoegiro, Dongdaemun-gu, Seoul 02455, Republic of Korea}

\author[0000-0001-5510-2803]{Thomas R. Quinn}
\affil{Department of Astronomy, University of Washington, Seattle, WA 98195, USA}

\author{the {\it AGORA} Collaboration}



\begin{abstract}

Given their highly nonlinear dynamics and sensitivity to initial conditions, galaxy mergers are a compelling area to conduct a simulation code comparison. We perform a comparative study of a major galaxy merger at $z \approx 4.5$ in cosmological zoom-in hydrodynamic simulations of a Milky Way-mass galaxy progenitor. The comparison employs the AGORA \texttt{CosmoRun} suite of nine well-calibrated, state-of-the-art numerical codes, each adopting a different stellar feedback scheme. We find that the evolution of the star formation rate (SFR) during the interaction is strongly shaped by the stellar feedback type. Using kinetic feedback in the feedback model drives a pronounced merger-induced starburst that starts to subside before coalescence; using thermal feedback without kinetic feedback yields prolonged SFR growth even after coalescence; and using delayed cooling or radiation pressure results in highly fluctuating SFR. Tracking gas particles in particle-based codes reveals that kinetic feedback facilitates gas inflow from the secondary galaxy onto the primary galaxy between the first periapsis and apoapsis, thus producing an earlier and more prominent starburst. In contrast, thermal feedback, augmented by superbubble or delayed-cooling feedback, suppresses gas cooling, creates a more extended gas distribution, and hinders strong starbursts during the merger. We also observe an inverse correlation between burst fraction and pre-merger gas fraction that is independent of feedback models. Overall, these results highlight the sensitivity of simulated galaxy mergers' star formation response to stellar feedback prescriptions. This study indicates that galaxy mergers may serve as a good testbed for stellar feedback processes in cosmological simulations.

\end{abstract}



\section{Introduction} \label{sec:intro}

Galaxy mergers are important processes in the hierarchical structure formation of the universe \citep{White+1978,Cole+2000,Springel+2006}. Recent observational studies show that galaxy mergers can lead to star formation rate (SFR) enhancement that continues for up to 1 Gyr after coalescence \citep{Ferreira+2025}, and post-merger galaxies exhibit a much more intense central starburst than isolated galaxies with similar global star-formation enhancements \citep{Thorp+2024}. Moreover, the gas metallicity gradients typically flatten shortly after the first pericenter passage of a merger \citep{Pan+2025}, and there is notable AGN enhancement in galaxy pairs at high redshift \citep{Duan+2026}. Using simulations, theoretical studies were also conducted to better understand the gas and stellar dynamics during a merger and to supply known values of merger parameters such as the mass ratio, gas mass, and orbital orientation, which are more difficult to measure observationally. Most theoretical studies of galaxy mergers have employed idealised, binary-system simulations, which enable easier systematic control of a merger's initial state and feedback prescriptions. Idealised simulations allow a more direct investigation of the causal dependence of key physical processes — including SFR enhancement \citep{Cox+2008}, metallicity evolution \citep{Rupke+2010}, bulge formation \citep{Hopkins+2009}, AGN activity \citep{Park+2017}, and tidal disruption event rates \citep{Li+2019} — on fundamental merger parameters such as mass ratio, progenitor stellar masses, gas mass fractions, and orbital configurations, as well as on the simulation feedback scheme.
However, these studies often do not include accretion from the cosmic web, may not fully account for the assembly history that shaped the progenitor galaxies, and often rely on manually specified orbital parameters, which may not always reflect realistic distributions from cosmology. Studies of mergers within fully cosmological, hydrodynamical simulations have also recently emerged to address those disadvantages. These studies typically employ public datasets of large-volume simulations developed by big collaborations, such as IllustrisTNG \citep{Patton+2020, Hani+2020}, SIMBA \citep{RodriguezMontero+2019}, FIRE \citep{Angles-Alcazar+2017}, EAGLE \citep{Qu+2017}, or Horizon-AGN \citep{Kaviraj+2015}. Even though these cosmological simulations allow investigations of galaxy mergers in a more realistic context, it is much more challenging to adjust variables to sample a complete parameter space in a merger's initial state, or to change the adopted feedback model because of the high computational cost. 

Comparisons of galaxy merger statistics across large cosmological simulations reveal substantial discrepancies. For instance, \cite{Patton+2020} reported that TNG100-1 exhibits higher specific star formation rate (sSFR) enhancement in interacting galaxy pairs compared to EAGLE and Illustris-1, and \cite{Quai+2023} also showed that TNG100-1 has the highest quenched fraction for post-merger galaxies. A key challenge when interpreting comparative studies of galaxy mergers in cosmological simulations is disentangling the interplay between the underlying numerical implementations of each simulation and intrinsic merger parameters, since the resulting properties of the merger remnant (the single system after two galaxies merge and coalesce) reflect both contributions. 
Regarding the numerical methodology, for example, the Illustris-1, TNG100-1, and EAGLE simulations differ in several notable respects, including the observables against which each simulation is calibrated, the adopted star formation prescriptions, the implementation of feedback processes, the assumed cosmological parameters, and the treatment of gas cooling/heating and the UV background \citep{Vogelsberger+2013, Genel+2014, Pillepich+2018, Schaye+2015}. Galaxy mergers are also intricate processes that are highly sensitive to many attributes of the progenitor galaxies (such as stellar mass, mass ratio, gas mass fraction, bulge-to-disc ratio, trajectory orientation, and progenitors' morphology; \citealt{Cox+2008, Hopkins+2009, Renaud+2022}), making the disentanglement even more difficult. Hence, when comparing results of galaxy mergers in these simulations, it is challenging to attribute the disagreements to a specific cause, as there can be multiple confounding factors.
A carefully calibrated suite of cosmological simulations is therefore crucial to unravel different underlying factors (progenitor galaxies' properties, feedback models, star formation criteria, etc .) contributing to a merger's outcomes.

Established in 2012, the AGORA (Assembling Galaxies of Resolved Anatomy) code comparison project has aimed at advancing the predictive capabilities of galaxy simulations and fostering our insight into the feedback mechanisms that regulate galaxy evolution. To this end, cross-comparisons are conducted between contemporary galaxy simulation codes that share various simulation parameters while differing only in the code architecture and the stellar feedback subgrid models. The comparisons enable discrepancies between codes to be explained by fewer degrees of freedom and allow results to reflect the numerical perspectives used in different simulation communities rather than code-dependent artifacts. This strength of AGORA makes it an ideal framework for comparing the effects of galaxy mergers in different simulation codes. The AGORA project began with the introductory proof-of-concept, dark-matter-only simulations using cosmological zoom-in initial conditions~\citep[hereafter Paper~\citetalias{Kim+2013}]{Kim+2013}, followed by a comparison of idealised, isolated Milky Way-mass galaxy simulations~\citep[hereafter Paper~\citetalias{Kim+2016}]{Kim+2016}. These foundational, crucial efforts laid the groundwork to develop \texttt{CosmoRun}, a fully cosmological zoom-in hydrodynamic simulation suite with eight numerical codes well-calibrated to each other~\citep[hereafter Papers~\citetalias{Roca-Fabrega+2021} and \citetalias{Roca-Fabrega+2024}]{Roca-Fabrega+2021, Roca-Fabrega+2024}. With \texttt{CosmoRun}, comparisons on different astrophysical questions have been studied, including the properties of satellite galaxies~\citep[hereafter Paper~\citetalias{Jung+2024}]{Jung+2024}, the properties of the circumgalactic medium~\citep[hereafter Paper~\citetalias{Strawn+2024}]{Strawn+2024}, the mechanisms of satellite quenching~\citep[hereafter Paper~\citetalias{Rodriguez-Cardoso+2025a}]{Rodriguez-Cardoso+2025a}, the formation and evolution of galactic disc~\citep[hereafter Paper~\citetalias{Jung+2025a}]{Jung+2025a}, and, most recently, the dark matter halo morphology ~\citep[hereafter Paper~\citetalias{Barrow+2026a}]{Barrow+2026a}. Starting from Paper~\citetalias{Jung+2025a}, the code GADGET-4 is added to the suite, expanding the total number of codes in \texttt{CosmoRun} to nine. Following the \texttt{CosmoRun} framework with adjustments to resolution and initial conditions, the AGORA collaboration also embarked on a new direction by targeting massive high-redshift galaxies ($\text{M}_\text{halo}=10^{10}\text{--}10^{11} \text{M}_\odot$ at $z = 10$)~\citep[hereafter Paper~\citetalias{Kim+2025}]{Kim+2025} to help explain JWST's discoveries of early massive luminous galaxies \citep[see review paper by][]{Adamo+2025}. Even though the participating codes in AGORA were carefully calibrated and identical in initial conditions, these previous AGORA papers still reported notable discrepancies in numerous astrophysical properties. 

In this paper, we continue this comparison endeavor by investigating how different code architectures and stellar feedback models in the \texttt{CosmoRun} simulation suite affect the outcome of a major galaxy merger. Paper~\citetalias{Roca-Fabrega+2024} showed that up to $z = 2$, the main galaxy (the zoomed-in galaxy in the simulation) in all participating codes experiences one major merger at $z \approx 4.5$ (mass ratio >~0.25) and two minor mergers at $z \approx 2.2$ (0.25~> mass ratio >~0.1). Despite having these three mergers, the present paper heavily focuses on the first one. To avoid confusion and repetition, we refer to it in the paper as \textit{the \textit{target} merger}. There are three reasons why only the first major merger is chosen for the comparison. Firstly, since the simulations are calibrated only up to $z = 4$ (Paper~\citetalias{Roca-Fabrega+2021}) and the \textit{target} merger happens shortly before that redshift, the merger's initial state is expected to converge between codes. Secondly, early mergers offer the best opportunity to compare the influence of subgrid prescriptions between codes. This is because numerical artifacts (for example, the timing discrepancy pointed out in Paper~\citetalias{Roca-Fabrega+2024}) and differences in other areas (such as satellite population, shown in Paper~\citetalias{Rodriguez-Cardoso+2025a}, or morphology, shown in Paper~\citetalias{Jung+2025a}) can compound over cosmological timescales and may overwhelm the comparison of later mergers. Thirdly, the timings of the two minor mergers at $z \approx 2.2$ overlap significantly. Specifically, the first periapsis of the second minor merger precedes the coalescence of the first minor merger. This temporal overlap makes disentangling the effects of individual mergers much more challenging. In addition, the snapshots of GIZMO and RAMSES do not extend to the end of the second minor merger, thereby limiting comparisons across all nine codes. Because of those reasons, this paper only centres on the first major merger at $z \approx 4.5$ in \texttt{CosmoRun}. This work constitutes part~1 (Paper IX - Part 1) of a two-part series dedicated to the analysis of this merger event. Specifically, this paper investigates the impact of the merger on star formation, while the companion part-2 paper (Nguyen et al. 2026, submitted, hereafter Paper IX - Part 2) examines the impact on stellar and gas morphology.

This paper is organised as follows. In Section~\ref{sec:method}, we describe the technical aspects of the AGORA \texttt{CosmoRun} simulation suite, the merger tree, the progenitors' initial properties, and our method to define merger timings and stages. Section~\ref{sec:effect_star_formation} analyses the effect of the galaxy merger on star formation, including identifying the SFR evolution pattern due to different stellar feedback schemes (Section~\ref{subsect:sfr_evolution_pattern}), analysing the gas properties during the interaction (Section~\ref{subsect:gas_properties_analysis}), and computing the burst fraction (Section~\ref{subsect:burst_fraction}). Section~\ref{sec:Discussion} discusses the comparison of our results with observational studies as well as other theoretical works and assesses the study's caveats. Finally, in Section~\ref{sec:conclusion}, we summarise the main findings of the paper.

\section{Methods} \label{sec:method}

\subsection{The AGORA \texttt{CosmoRun} Simulation Suite}

AGORA \texttt{CosmoRun} is a suite of high-resolution cosmological zoom-in hydrodynamic simulations of a Milky-Way–mass halo ($\text{M}_\text{halo} \approx 10^{12}\;\text{M}_\odot$ at $z = 0$) run with multiple cosmological simulation codes. All simulations assume a flat $\Lambda$CDM cosmology and were run with the cosmological parameters obtained from the WMAP7/9+SNe+BAO results~\citep{Komatsu+2011, Hinshaw+2013}: $\Omega_{m} = 0.272$, $\Omega_{\Lambda} = 0.728$, $h = 0.702$, $\sigma_{8} = 0.807$, and $n_{s} = 0.961$. All simulations use identical cosmological initial conditions generated with \textsc{music} \citep{Hahn+2011}, starting at $z = 100$ and reaching $z \leq 2$ (with some codes reaching $z \approx 0$ \footnote{The decision of running down to lower redshifts was made by each code group and is not indicative of the computational cost or the code performance.}). The current \texttt{CosmoRun} suite consists of nine codes: three adaptive mesh refinement (AMR) codes, ART-I~\citep{Kravtsov+1997}, ENZO~\citep{Bryan+2014}, and RAMSES~\citep{Teyssier+2002}; four particle-based codes, CHANGA~\citep{Jetley+2008a, Jetley+2010, Menon+2015}, GADGET-3 (GADGET3-OSAKA version; \citealp{Shimizu+2019, Nagamine+2021}; base code by \citealp{Springel+2005}), GADGET-4 (GADGET4-OSAKA version; \citealp{Romano+2022, Romano+2022a, Oku+2024, Granizo+2026}; base code by \citealp{Springel+2021}), and GEAR (\citealp{Revaz+2012}; base code by \citealp{Springel+2005}); one moving-mesh code, AREPO~\citep{Springel+2010, Weinberger+2020}; and one meshless/hybrid code, GIZMO~\citep{Hopkins+2015}. We refer to the version of AREPO used in this study as AREPO-T, which represents the AREPO code run with only thermal feedback. 

Regarding the \textsc{music} initial conditions, all simulations in the suite have the total comoving volume of~60 $\text{(Mpc/h)}^{3}$. We used a~$128^{3}$ root resolution and created a zoom-in region using five additional levels of nested refinement to reach an effective resolution of~$4096^{3}$ in the highest-resolution region. 
The three AMR codes in \texttt{CosmoRun} have the finest cell size of~163 comoving pc ($\approx 54$ physical pc at $z=2$), achieved through seven additional refinement levels in the zoom-in region. Cells are refined if the baryon mass or the particle mass in the cell exceeds four times the mean density of the subgrid. ENZO also uses \texttt{MustRefineParticles} that make cells around them refined at least down to~20.9 comoving kpc. The other six codes (using particle-based, moving mesh, or meshless methods) employ a gravitational force softening length of~800 comoving pc within the zoom-in region for $z > 9$ and of~80 proper pc thereafter. Inside the zoom-in region, the most refined dark matter (DM) particles have a mass of $m_{\text{DM}, \text{IC}}=2.8\times10^{5}\msun$ and the most refined gas particles have a mass of $m_{\text{gas}, \text{IC}}=5.65\times10^{4}\msun$ (for codes that use them). For more details about the initial conditions, we refer readers to Paper~\citetalias{Kim+2013} and Paper~\citetalias{Roca-Fabrega+2021}. 

To ensure a robust comparison, all \texttt{CosmoRun} simulations share many common astrophysical packages, including the radiative gas cooling library \textsc{grackle} \citep{Smith+2017}, the redshift-dependent cosmic UV background \citep{Haardt+2012}, and the star formation criteria. The gas density threshold for star formation is $n_\text{H,thres}=1\, \text{cm}^{-3}$, where $n_\text{H}$ is the total hydrogen number density (equivalent to $\rho_\text{gas,thres} \approx 2.2\times10^{-24} \text{g}\,\text{cm}^{-3}$, assuming a primordial hydrogen fraction of~0.752). Star particles can be created from gas elements satisfying the density threshold at a rate of $d\rho_{\star}/dt = \epsilon_\star\rho_\text{gas}/t_{\text{ff}}$, where $\epsilon = 0.01$ is the formation efficiency and $t_\text{ff}=\sqrt{3\pi/32G\rho_{\text{gas}}}$ is the local freefall time. Each code group is free to choose whether this process follows a stochastic or deterministic nature. 
For particle-based codes and GIZMO, a star particle inherits the mass of its parent gas particle (except CHANGA, which uses a constant initial stellar mass of $5.65\times10^{4}\msun$). 

\begin{table*}
\vspace*{1mm}
\caption{\footnotesize Stellar feedback implementation adopted by each code group. Adapted and synthesized from Papers \citetalias{Roca-Fabrega+2021}, \citetalias{Roca-Fabrega+2024}, and \citetalias{Jung+2025a}.  \tablenotemark{\textdagger}}
\centering
\begin{tabular}{c  c  c  c }
\hline\hline 
Code & Stellar feedback &  Runtime parameters\\ 
\hline
{\sc Art-I} & T+K, RP  & $E_{\rm thermal}= 2\times 10^{51}\,{\rm ergs/SN}$,\, $p =3.6 \times 10^6 \msun \, {\rm km \, s^{-1}/SN}$  \\
{\sc Enzo} & T  & $E_{\rm thermal} = 5\times 10^{52}\,{\rm ergs/SN}$ \\ 
{\sc Ramses} & T, DC  & $E_{\rm thermal}= 4\times 10^{51}\,{\rm ergs/SN}$, $\sigma_{\rm min}=100\,\,\rm{km \, s^{-1}}$, $\,\,T_{\rm delay}=10\,\,\rm{Myr}$  \\
{\sc Changa} & T+S  & $E_{\rm thermal} = 5\times10^{51}\,{\rm ergs/SN}$  \\  
{\sc Gadget-3} & T+K, RP, DC   & $E_{\rm SN} = 4\times 10^{49}\,{\rm ergs/ \msun} (\rm T+K), E_{\rm ESFB} = 6\times 10^{49}\,{\rm ergs/ \msun} (\rm only \, T), \,\,\,T_{\rm delay}=t_{\,\rm hot}$ \\
{\sc Gadget-4} & T+K  & $E_{\rm SN} = 1\times 10^{52}\,{\rm ergs/SN}$ \\ 
{\sc Gear} & T, DC  & $E_{\rm thermal} = 4.5\times10^{51}\,{\rm ergs/SN}$, $\,\,T_{\rm delay}=5\,\,\rm{Myr}$  \\
{\sc Arepo-T} & T  & $E_{\rm thermal} = 2\times10^{52}\,{\rm ergs/SN}$ released 3 Myr after star formation  \\
{\sc Gizmo} & T+K & $E_{\rm SN} = 5\times10^{51}\,{\rm ergs/SN}$  \\  
\hline 
\end{tabular}
\tablenotetext{$\textdagger$}{\scriptsize T = thermal feedback, K = kinetic feedback, RP = radiation pressure, DC = delayed cooling, S = superbubble. ESFB = early stellar feedback. We note that the literature sometimes distinguishes between kinetic and mechanical feedback \citep{Rosdahl+2017}. Kinetic feedback injects momentum set by a free parameter, whereas mechanical feedback injects momentum according to whether the Sedov--Taylor phase of the supernova (SN) is resolved. In \texttt{CosmoRun}, ART-I and GADGET-3 adopt the former, while GADGET-4 and GIZMO adopt the latter. Because both approaches are based on injecting momentum, for simplicity, we refer to them collectively as "kinetic feedback" (K). The feedback scheme of each code group follows the prevalent practice in its community as closely as possible. For more details about the specific implementation, the readers can refer to Papers \citetalias{Roca-Fabrega+2021}, \citetalias{Roca-Fabrega+2024}, and \citetalias{Jung+2025a} and the references within.}
\label{tab:feedback}
\vspace*{2mm}
\end{table*}

Although \texttt{CosmoRun} simulations share many common baryonic physics, one of the key differences between the codes is the stellar feedback prescriptions, which is also the focus of this comparison. Each code group sets its stellar feedback prescription as close to the widely adopted one in its own code community as possible, while striving to achieve the main halo's stellar mass at $z=4$ that accords with semi-empirical model predictions. At $z=4$ (near the end of the first major merger), all the codes converge on the stellar and total masses within half an order of magnitude of one another. A quick summary of the codes' stellar feedback implementations that are relevant to this paper is shown in Table~\ref{tab:feedback}. Because \texttt{CosmoRun} does not include AGN feedback, for brevity, all subsequent mentions of "feedback" in the paper refer exclusively to stellar feedback. Readers may check Paper~\citetalias{Roca-Fabrega+2021} as well as the appendix of Papers~\citetalias{Roca-Fabrega+2024} and \citetalias{Jung+2025a} for more elaboration on the listed feedback models. 

As emphasised in Paper~\citetalias{Strawn+2024}, it is important to note that the participating codes in AGORA do not necessarily capture the full diversity of their broader code communities, since each community can still have variations in its own stellar feedback prescriptions and user-defined parameters. Thus, studies employing AGORA codes with non-\texttt{CosmoRun} feedback models should be cautious when drawing comparisons between their results and the corresponding \texttt{CosmoRun} results.

\subsection{Halo Finding and Stellar Assignment}
\label{subsect:halofinding}

We identified DM halos and the merger histories using \textsc{HASKAP PIE} \citep[][Paper~\citetalias{Barrow+2026a}]{Barrow+2026}, an all-in-one algorithm that both finds halos and builds merger trees using overdensity finding, energy solving, cluster finding, and particle tracking. A key feature of \textsc{HASKAP PIE} is its energy-based approach, which allows it to define halos in a non-spherical fashion using convex hulls and ensures that only mutually bound DM particles are included in the halo structure. All halos found by \textsc{HASKAP PIE} contain only refined DM particles. The algorithm is also capable of assigning stars to DM halos. Each star particle's orbital energy is calculated with respect to all DM halos, and a star is assigned to halos with which it has negative orbital energy. This energy-based approach allows our stellar assignment to be robust during halo interactions. 

\begin{figure}
    \centering
    \includegraphics[width=0.95\linewidth]{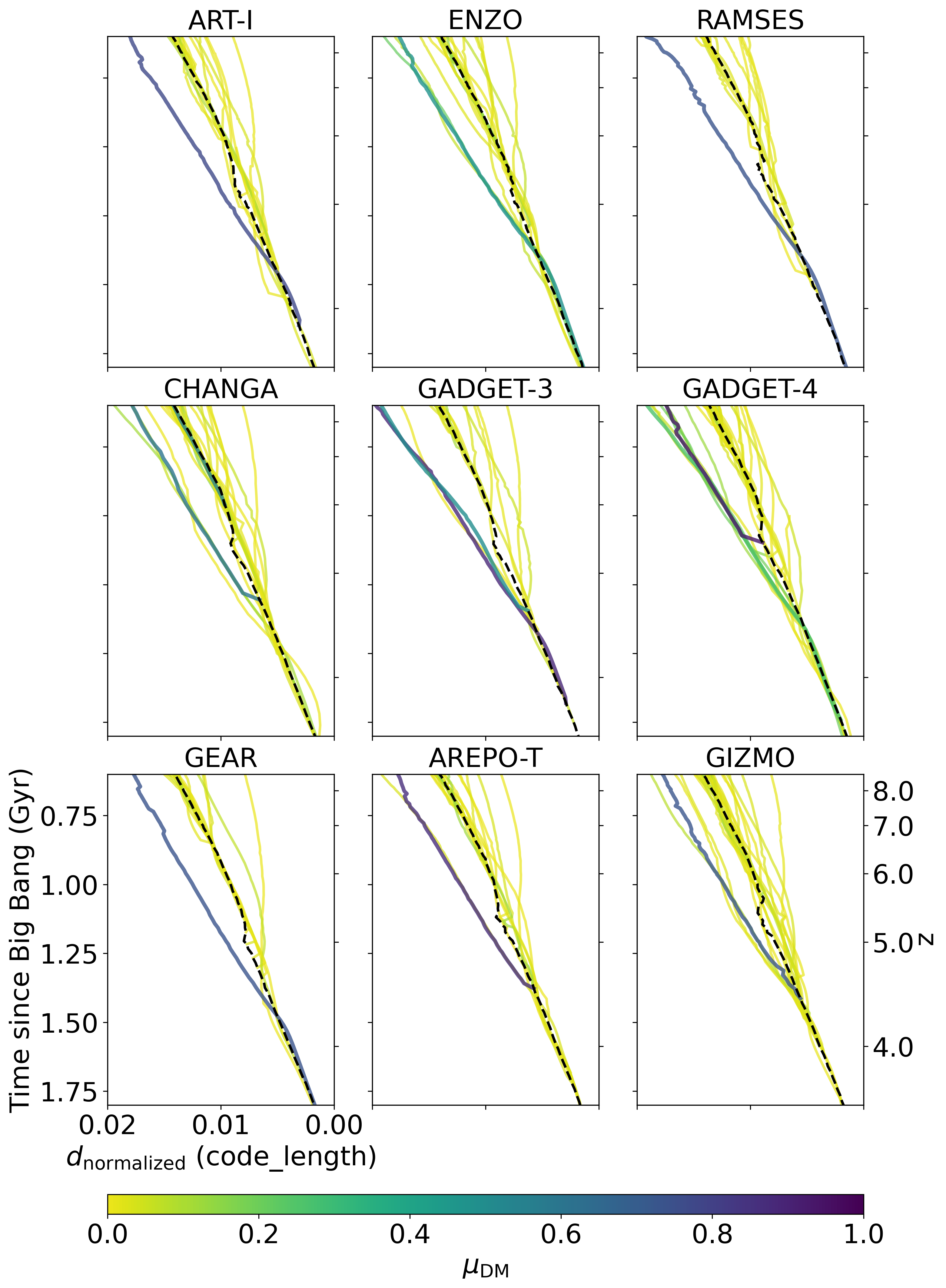}
    \caption{The merger history of the main halo (dashed black line) when the \textit{target} merger happens at $z \approx 4.5$. The subhalos are shown by solid lines and coloured by their DM mass ratio with respect to the main halo. For clearer visualization, we only plot subhalos with $\mu_\text{DM} > 0.005$ that form outside the main halo and then merge with it within the plotting period. The \textit{target} merger ($\mu_\text{DM} > 0.45$) clearly appears in all nine codes. As other surrounding mergers are very minor, we attribute any changes in galaxy properties during this period primarily to the \textit{target} merger.}
    \label{fig:merger_tree}
\end{figure}

Fig.~\ref{fig:merger_tree} shows the \textsc{HASKAP PIE} merger tree of the main halo (dashed black line) during the \textit{target} merger ($z \approx 4.5$). The distance on the x-axis is calculated relative to the centre of the main halo at two Gyr after the Big Bang, which was selected as a reference point to visualize the merger tree. The line colour represents the DM mass ratio ($\mu_\text{DM}$) between the sub-halo and the main halo, evaluated immediately before the first overlap of the two halos' convex hulls.

We can see the \textit{target} merger ($\mu > 0.45$) clearly in all nine participating codes. In the period surrounding and preceding this merger, the remaining interactions are comparatively minor ($\mu < 0.05$). Therefore, we assume that any changes to star formation and morphology of the main galaxy during this period are attributed primarily to the \textit{target} merger. We discuss a caveat stemming from this assumption in Section~\ref{sec:caveat}. In the remainder of the paper, we refer to the higher-mass halo/galaxy in the merging interaction as the primary halo/galaxy and the lower-mass halo/galaxy as the secondary halo/galaxy. In \texttt{CosmoRun}, the main halo (the most massive halo in the zoom-in region) is always the primary halo in all of its mergers. 

In a major merger, the non-negligible mass of the secondary halo affects how \textsc{HASKAP PIE} identifies the main halo. Because \textsc{HASKAP PIE} defines a halo based on gravitationally bound particles, the main halo's centre of gravity can be shifted outside of its densest core if a considerable DM mass from the sub-halo becomes bound to the main halo during the infall. 
This shift in the centre of gravity explains the rapid change in the path of the main branch in Fig.~\ref{fig:merger_tree} (dashed line). 
Another feature in the merger tree is the existence of subhalos within subhalos. With its accuracy, \textsc{HASKAP PIE} allows us to detect and track subhalos of the secondary halo. For example, in ENZO, GADGET-3, and GADGET-4, we notice several halos with $\mu_\text{DM} > 0.2$ coming in the same direction as the secondary halo. These halos are not isolated halos but reside in a DM halo cluster and share bound DM particles with the secondary halo (Paper~\citetalias{Barrow+2026a}). 
Nevertheless, even when we take into account the subhalos of the secondary halo, the DM halos' merging time of the \textit{target} merger does not fully agree among the codes. 
This timing disagreement will be explored in a future paper with \textsc{HASKAP PIE} being run on the dark-matter-only counterpart simulations of \texttt{CosmoRun}. Besides the mentioned complications, the DM halos' interacting gravitational potentials are on a much larger scale than the interacting galaxies, while we are mainly interested in the changes in the stellar and baryonic components. Therefore, it would be more helpful to define the merger timings using the stellar components rather than the DM components. 

\subsection{Merger trajectory}
\label{subsect:merger_trajectory}

\begin{figure}
    \centering
    \includegraphics[width=\linewidth]{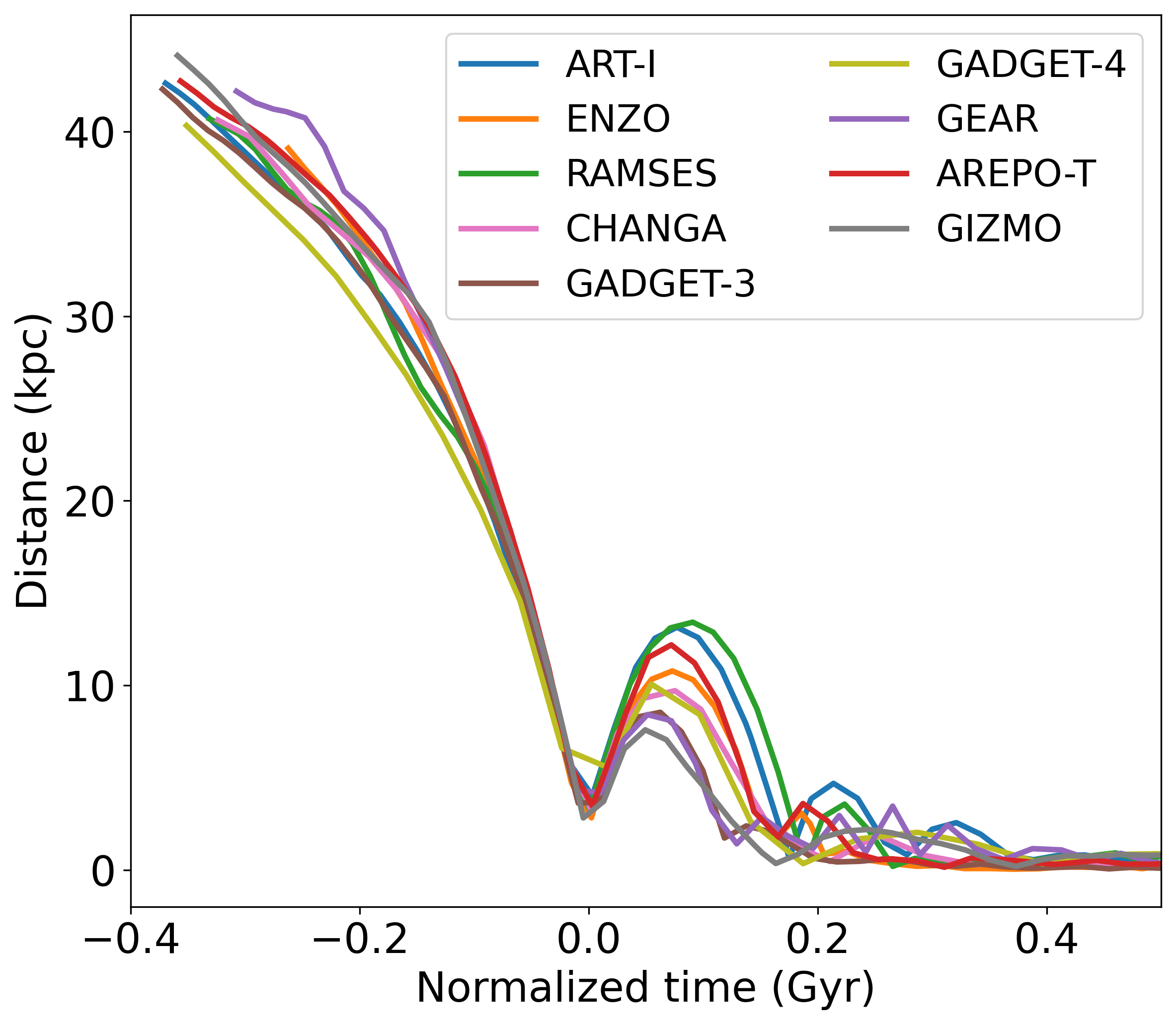}
    \caption{The orbital trajectory of the secondary galaxy relative to the primary during the \textit{target} merger. The time axis is set to zero at the time of the first periapsis. The distance is computed by using the top 10\%-bound star particles (stellar cores) from both galaxies at the beginning of infall. Notably for this merger, mesh-based codes predict a larger first-apoapsis distance compared to particle-based codes.}
    \label{fig:merger_distance}
\end{figure}

To track the distance between the two galaxies during the interaction, we identified the bound star particles of each galaxy from \textsc{HASKAP PIE} at the beginning of the infall. We excluded the particles that are beyond~1.5 times the interquartile range around the median in each position and velocity axis, and then selected the~10\% most bound star particles. These two subsets of star particles represent the two stellar cores, whose constituent stars should be close in both their position and velocity. Each stellar core comprises at least~250 star particles after the selection. We also define the galactic centre of each galaxy as the centre of mass of its stellar core. We tracked the locations of the star particles in the stellar cores over time using their particle IDs. 

Fig.~\ref{fig:merger_distance} displays the orbital trajectory of the interaction by tracking the distance between the two stellar cores. 
We notice a systematic difference in the first apoapsis distance, where all mesh-based codes (ART-I, ENZO, RAMSES, and the moving-mesh code AREPO-T) show a larger distance compared to all particle-based codes and the mesh-free code GIZMO. Specifically, the apoapsis distance of the mesh-based codes ($\approx 12\text{--}14$ kpc) is about~1.5 times larger than that of the particle-based codes ($\approx 8\text{--}9$ kpc). Nonetheless, examination of the trajectories of other mergers does not reveal a comparably clear trend. Thus, future work is needed to further explore whether the choice of code architecture systematically biases the apoapsis distance of a merger. 

\subsection{Merger timing and stages}
\label{subsect:merger_timing_and_stages}

\begin{figure*}
	\gridline{\fig{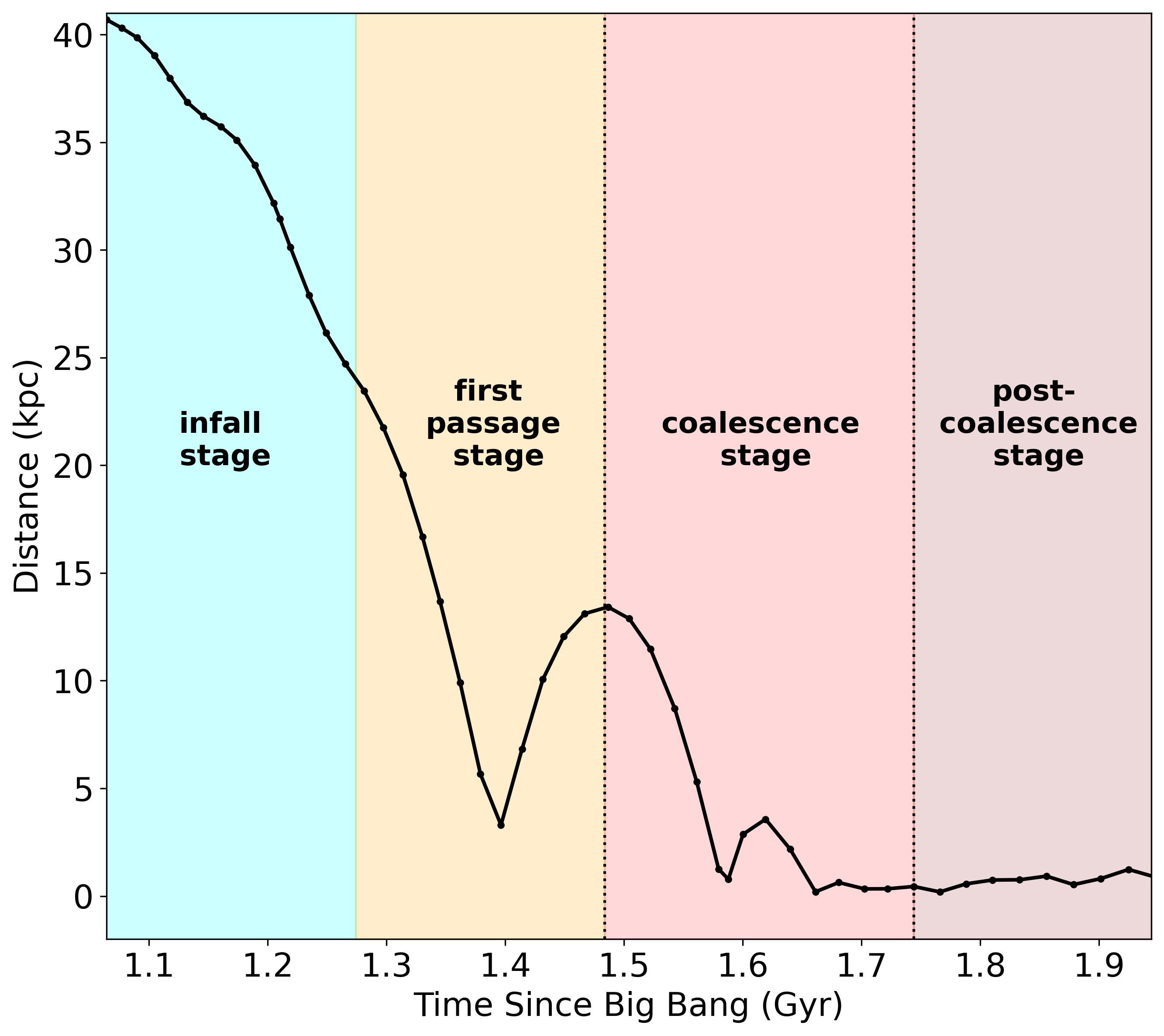}{0.45\textwidth}{}
    \fig{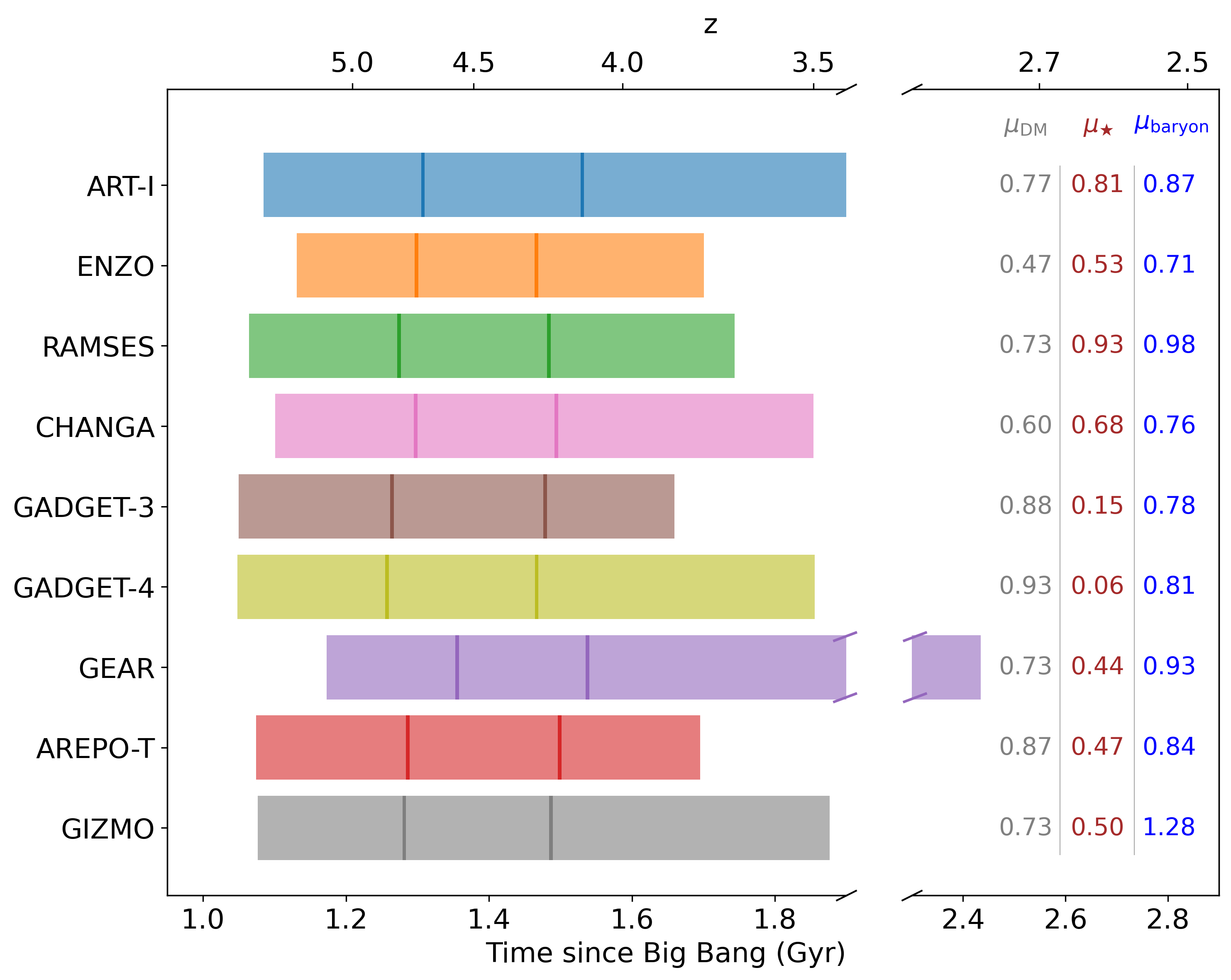}{0.54\textwidth}{}
		}
    \vspace{-20pt}
	\caption{Left: The four merger stages in our analysis: infall stage, passage stage, coalescence stage, and post-coalescence stage (or post-merger stage). The orbital trajectory of RAMSES's \textit{target} merger is used as an example. Right: The duration of the \textit{target} merger in each code, from the beginning of the infall stage to the end of the coalescence stage. The vertical coloured lines separate the merger timescale into the infall, first passage, and coalescence stages. The DM mass ratio $\mu_\text{DM}$ (in grey), stellar mass ratio $\mu_\bigstar$ (in red), and baryonic (star plus cool gas) mass ratio $\mu_\text{baryon}$ (in blue) of the merger are listed next to each code's bar. There are noticeable differences in the duration and the mass ratios between the codes.}
	\label{fig:merger_timing_and_stages}
\end{figure*}

We define the timings of a merger as follows. The start of a merger ($t_\text{start}$) is set at the timestep when the convex hulls of the two non-spherical DM halos overlap for the first time. We also refer to the timestep preceding $t_\text{start}$ as the pre-infall timestep ($t_\text{pre-infall}$). The end of a merger occurs when the stellar cores of two halos coalesce. To determine the coalescence time ($t_\text{cls}$), Paper~\citetalias{Roca-Fabrega+2024} used a fixed time of $\approx 300\,\text{Myr}$ after the time when the centres of the two merging halos are first closer than~5 kpc. This definition, however, does not scale with the merger's mass ratio because smaller satellites often take a longer time to merge with the host at a given redshift and orbital circularity \citep{Boylan-Kolchin+2008}. Moreover, halos take a longer time to merge at lower redshift due to the free-fall time getting longer \citep{Barrow+2026}. Therefore, we employed a different definition that involves the distance and the relative velocity between the two merging galaxies. We define the coalescence conditions as,
\begin{align}
    d  < \max(0.025&R_{\text{200,pri}}, 10\epsilon), \nonumber \\
    v_\text{rel}/\sigma_{v, \text{pri}} & < 0.1, \label{eq:coalescence_condition} \\  
    v_\text{rel}/\sigma_{v, \text{sec}} & < 0.1, \nonumber
\end{align}
where $d$ is the distance between the centres of mass of the stellar cores of the two interacting galaxies, $R_{\text{200c,pri}}$ is the $R_{200c}$ of the primary galaxy's DM halo (the radius enclosing an average DM density~200 times the universe's critical density at a given redshift, note that \textsc{haskap pie} only includes bound particles and uses the centre of gravity rather than the centre of mass), $\epsilon$ is the softening length (for particle-based and hybrid codes) or the smallest grid size (for AMR code) in proper unit, $v_\text{rel}$ is the centre-of-mass velocity of the stellar core of the secondary galaxy relative to that of the primary galaxy, and $\sigma_{v} = \sqrt{\sigma_x^{2} +\sigma_y^{2}+\sigma_z^{2}}$ is the 3D velocity dispersion of the galaxy's stellar core ("pri" is for the primary galaxy and "sec" is for the secondary galaxy). We show the time evolution of each variable and further discuss our criteria in Appendix~\ref{appendix:evaluating_massratio_coalescence}.

To better isolate the physical effects at each phase of the \textit{target} merger, we divided the interaction into distinct stages. The stage division is adapted from \cite{Lotz+2008} with some modifications. First, we determined the starting timestep of the merger using \textsc{HASKAP PIE} ($t_{\mathrm{start}}$), the timestep of the first apoapsis ($t_{\mathrm{max}}$), and the timestep of coalescence ($t_{\mathrm{cls}}$, using the conditions in Equation~\ref{eq:coalescence_condition}). We then divided the merger into four stages: infall, first passage, coalescence, and post-coalescence (or post-merger), as shown in Table~\ref{tab:merger_stages} and the left subplot of Fig.~\ref{fig:merger_timing_and_stages}.

\begin{table}
\caption{The names and timings of our merger stages. $t_\text{start}$ is the timestep when the two non-spherical DM halos overlap for the first time. $t_\text{max}$ is the timestep of the first apoapsis. $t_\text{cls}$ is the coalescence timestep, determined using Equation~\ref{eq:coalescence_condition}. The colour codes are used in Figs.~\ref{fig:merger_timing_and_stages}, \ref{fig:mass_sfr}, and \ref{fig:CHANGA_SFR_delayed_starburst}.}
\centering
\footnotesize
\begin{tabular}{cccc}
\hline \hline

Stage & Starting  & Ending  & Colour \\
\hline

Infall & $t_{\mathrm{start}}$ &  $(t_{\mathrm{start}} + t_{\mathrm{max}})/2$  & \fcolorbox{black}{mycyan}{\rule{0pt}{4pt}\rule{4pt}{0pt}} \rule[-1ex]{0pt}{0pt} \\
First passage& $(t_{\mathrm{start}} + t_{\mathrm{max}})/2$ &  $t_{\mathrm{max}}$  & \fcolorbox{black}{myorange}{\rule{0pt}{4pt}\rule{4pt}{0pt}} \rule[-1ex]{0pt}{0pt} \\
Coalescence & $t_{\mathrm{max}}$ &  $t_{\mathrm{cls}}$  & \fcolorbox{black}{myred}{\rule{0pt}{4pt}\rule{4pt}{0pt}} \rule[-1ex]{0pt}{0pt} \\
Post-coalescence &  $t_{\mathrm{cls}}$ &  -  & \fcolorbox{black}{mydarkred}{\rule{0pt}{4pt}\rule{4pt}{0pt}} \rule[-1.8ex]{0pt}{0pt} \\
\hline    
\end{tabular}
\label{tab:merger_stages}
\end{table}

The timescales of the \textit{target} mergers (from $t_\text{start}$ to $t_\text{cls}$) in the nine \texttt{CosmoRun} codes are shown in the right subplot of Fig.~\ref{fig:merger_timing_and_stages}. Within each bar, the vertical colored lines mark the boundaries between the three stages of the merger timescale (infall, first passage, and coalescence). The right subplot of Fig.~\ref{fig:merger_timing_and_stages} displays the "timing discrepancy" of the merger among the codes (as also mentioned in Paper~\citetalias{Roca-Fabrega+2024}). This timing discrepancy is most prominent for GEAR, whose time of the first periapsis is offset by $60\,\rm{Myr}$ compared to the average time of the rest of the codes. It is important to note that we used cubic splines to interpolate the trajectory of the merger when the time spacing between snapshots fails to capture the exact moment of the first periapsis. This issue of timing discrepancy was investigated in Appendix C of Paper~\citetalias{Roca-Fabrega+2024}, where they concluded that the issue's origin in GEAR is the user parameters (\texttt{ErrTolIntAccuracy} and \texttt{MaxRMSDisplacementFac}).

Consistent with the method in Paper~\citetalias{Roca-Fabrega+2024}, the two progenitor galaxies in most codes reach coalescence about $\approx 300\,\text{Myr}$ after the first periapsis. The merger timescale is also comparable across most codes, being about $600\text{--}800\,\text{Myr}$. One exception is GEAR's \textit{target} merger, where the galaxies take almost $1.3\,\text{Gyr}$ to coalesce. Overall, our merger timescales are still consistent with the range of $0.5\text{--}2$ Gyr found by \citet{Lotz+2008} for simulated major mergers.

\subsection{Merger mass ratio and initial properties}

\begin{table*}
\vspace*{1mm}
\caption{The properties of the two progenitor galaxies at the pre-infall timestep and the orientation of the \textit{target} merger. From top to bottom: $z_\text{pre-infall}$ is the pre-infall redshift at
which the properties are reported; $M_\text{DM}$, $M_{\bigstar}$, and
$M_\text{gas}$ are the DM, stellar, and total gas masses, with subscripts "pri" and "sec" denoting the primary and secondary galaxy; and $\theta_\text{pri}$ ($\theta_\text{sec}$) is the angle between the merger's orbital angular momentum and the primary (secondary) galaxy's rotational angular momentum.}
\centering
\footnotesize
\begin{tabular}{c|ccccccccc}
\hline \hline
Code                    & ART-I   & ENZO    & RAMSES  & CHANGA  & GADGET-3 & GADGET-4 & GEAR    & AREPO-T & GIZMO   \\
\hline
$z_\text{pre-infall}$ & 5.53 & 5.33 & 5.59 & 5.49 & 5.65 & 5.70 & 5.18 & 5.55 & 5.54 \\
$\log_{10}(M_\text{DM,pri}/\msun)$   &  10.55 & 10.79 & 10.58 & 10.55 & 10.54 & 10.51 & 10.63 & 10.54 & 10.60 \\
$\log_{10}(M_\text{DM,sec}/\msun)$      & 10.44 & 10.46 & 10.45 & 10.33 & 10.48 & 10.48 & 10.50 & 10.47 & 10.46 \\
$\log_{10}(M_{\bigstar\text{,pri}}/\msun)$     & 8.78 & 8.37 & 8.55 & 8.33 & 8.85 & 8.88 & 8.82 & 8.17 & 8.08 \\
$\log_{10}(M_{\bigstar\text{,sec}}/\msun)$     & 8.68 & 8.10 & 8.52 & 8.16 & 8.03 & 7.67 & 8.46 & 7.84 & 7.77 \\
$\log_{10}(M_\text{gas,pri}/\msun)$       & 9.71 & 10.00 & 9.67 & 9.82 & 9.84 & 9.83 & 10.05 & 9.81 & 9.65 \\
$\log_{10}(M_\text{gas,sec}/\msun)$        & 9.63 & 9.86 & 9.68 & 9.67 & 9.79 & 9.78 & 9.95 & 9.72 & 9.78 \\
$\theta_\text{pri}$ ($\degree$)     & 126     & 27 & 55 & 120 & 146  &        96  & 130 & 98  & 38 \\
$\theta_\text{sec}$ ($\degree$)     &  26     & 19 & 8 & 44 & 67  &  13      & 25 & 7  & 133 \\
\hline    
\end{tabular}
\label{tab:initial_state}
\end{table*}

We evaluated the merger's mass ratios at $t_\text{pre-infall}$. In the right subplot of Fig.~\ref{fig:merger_timing_and_stages}, the merger's DM mass ratio (gray), stellar mass ratio (red), and baryonic mass ratio (blue; defined as the sum of stellar mass and gas mass with $T < 10^{4.5}$ K) are annotated alongside each code's bar. The definition of baryonic mass as stellar mass plus cold gas or star-forming gas is also used in previous studies of galaxy mergers \citep{Stewart+2009, Hopkins+2010, Rodriguez-Gomez+2015}. All codes agree that the \textit{target} merger is a major merger, as the DM mass ratios and the baryonic mass ratios are about $0.6\text{--}0.9$, which are all above the commonly adopted values (0.25, 0.3, or 1:3) to classify major mergers in theoretical studies \citep{Cox+2008, Stewart+2009, Hopkins+2010, Lotz+2010}. One exception is the baryonic mass ratio in GIZMO. Because the secondary galaxy contains more gas than the primary, this ratio exceeds one, but the relative ratio of~1/1.28=0.78 still classifies the merger as a major merger. We also verified that even though there is a relatively wide range of the DM mass ratio among the codes, if we allow uncertainties of $\approx 100\,\text{Myr}$ in determining the pre-infall timestep, the DM mass ratios of the mergers converge better. In contrast, the stellar mass ratios (red numbers in the right subplot of Fig.~\ref{fig:merger_timing_and_stages}) exhibit a more noticeable discrepancy among the codes, even after allowing uncertainties for the pre-infall timestep. The mergers in ART-I, RAMSES, and CHANGA have the largest stellar mass ratios ($> 0.65$), while ENZO, GEAR, AREPO-T, and GIZMO show the value of $\approx 0.5$. Most severely, in both GADGET-3 and GADGET-4, the stellar mass ratios are not even large enough to classify the merger as a major merger, especially in observational studies where only stellar mass is used to calculate a merger's mass ratio \citep{Ellison+2008, Casteels+2014, Duncan+2019}. Indeed, the inconsistency between using DM mass and stellar mass ratio in classifying mergers was investigated in \cite{Stewart+2009a} and \cite{Hopkins+2010}, where they showed that a major DM merger can be a minor stellar merger because the mapping between a halo's DM mass and stellar mass is not a simple linear relation \citep{Conroy+2009a} and is dependent on multiple galaxy formation processes. Nevertheless, we do not find that the small stellar mass ratios in GADGET-3 and GADGET-4 systematically affect the study's main results. Hence, for the rest of the paper, a major merger is defined using the DM mass ratio. 

Table~\ref{tab:initial_state} shows the different types of mass (DM, stellar, gas) of the two progenitor galaxies at $t_\text{pre-infall}$. The DM and total gas masses are consistent across all codes to within~0.3–0.4 dex, whereas the stellar mass exhibits a larger difference of about~0.8 dex. Nevertheless, this difference in stellar mass remains broadly in agreement with the \texttt{CosmoRun} stellar mass calibration, which requires all codes to reach $1\text{--}5\times10^{9} \msun$ (0.7 dex range) at $z = 4$, as motivated by semi-empirical models (fig.~12 of Paper~\citetalias{Roca-Fabrega+2021}; fig.~4 of ~\citetalias{Roca-Fabrega+2024}). The initial relative velocity between the two galaxies also agrees well across the codes, as shown in Fig.~\ref{fig:d_v_j} in Appendix~\ref{appendix:evaluating_massratio_coalescence}. The orbital orientations, quantified by the angles between the progenitor galaxies' rotational axes and the merger's orbital axis, show the most variability between codes (last two rows of Table~\ref{tab:initial_state}). The rotational axis is computed by summing the angular momentum of all gravitationally bound stars within the halo's convex hull. The orbital angular momentum is calculated using the relative position and velocity vectors of the secondary galaxy with respect to the primary galaxy at $t_\text{pre-infall}$. The values of $\theta_\text{pri}$ and $\theta_\text{sec}$ indicate that both progenitors can have tilted prograde ($0^{\degree} \leq \theta < 90^{\degree}$), polar ($\theta \approx 90^{\degree}$), or tilted retrograde ($90^{\degree} < \theta \leq 180^{\degree}$) motion during the interaction. Nonetheless, most of our progenitor galaxies are dispersion-dominated and have very low disc-to-total ratios (Paper~IX - Part~2). Thus, the angular momentum direction is subjected to greater variability among codes due to the randomness in stellar motions. 

In summary, the initial physical properties are broadly consistent across the codes, providing a common baseline against which the merger-induced star formation response can be compared.

\section{Results}
\label{sec:effect_star_formation}

Previous studies from both simulations \citep{Lotz+2008, Lotz+2010} and observations \citep{Ferreira+2025} have shown that star formation enhancement usually peaks at a particular moment during the merger sequence. The merger stages defined in Table~\ref{tab:merger_stages} help us detect systematic patterns and better identify differences among the codes with regard to star formation (Section~\ref{subsect:sfr_evolution_pattern}). Once such discrepancies are established, we show in Section~\ref{subsect:gas_properties_analysis} that gas properties and gas particle tracking can provide a direct diagnostic for interpreting their cause.

\subsection{SFR evolution pattern during the major merger}
\label{subsect:sfr_evolution_pattern}

\begin{figure*}[tbh]
    \centering
    \includegraphics[width=\linewidth]{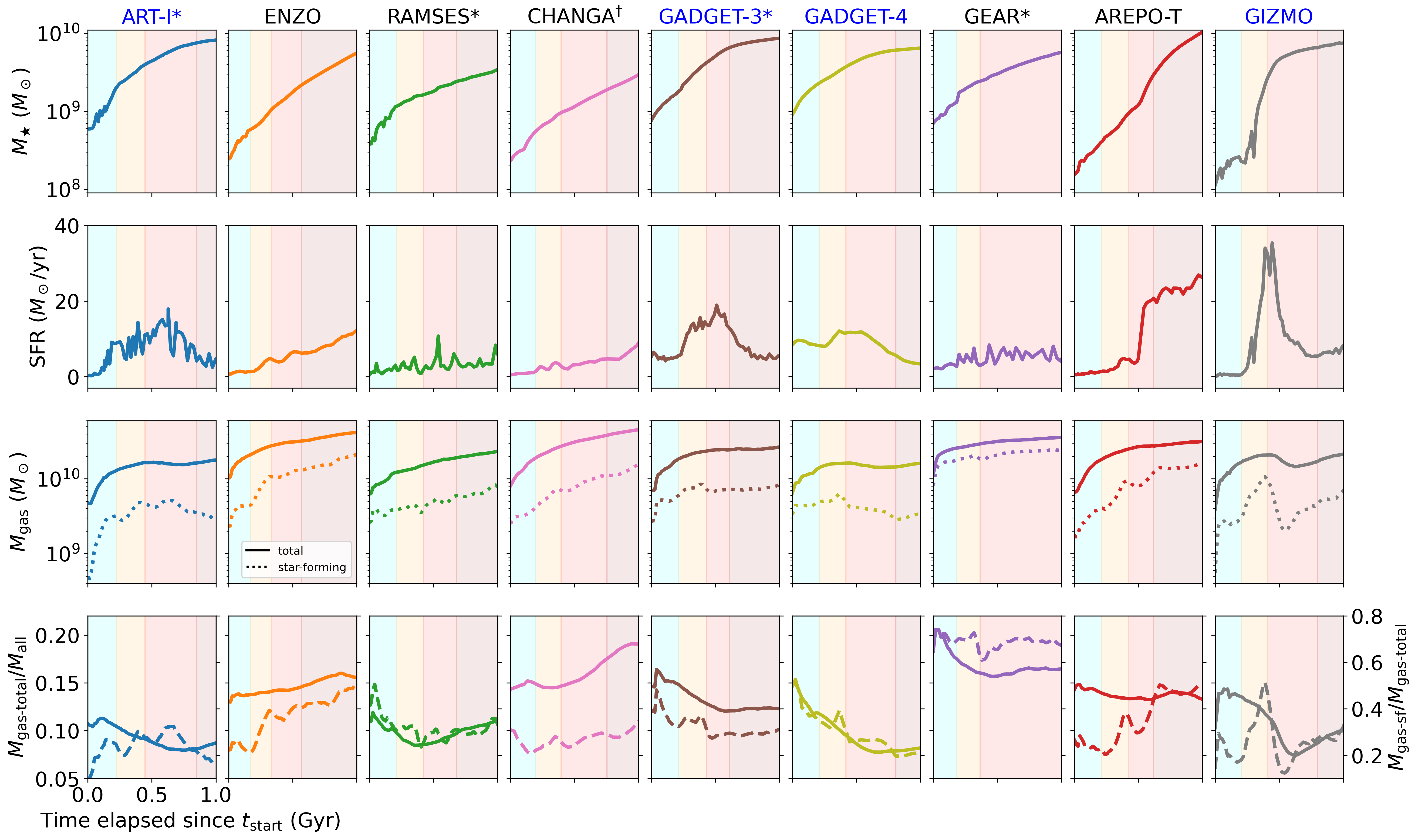}
    \caption{The evolution of the primary galaxy's stellar mass, SFR, gas mass, star-forming gas mass ($n_\text{H} >  1\,\text{cm}^{-3}$), and gas mass fraction, within~1 Gyr after the beginning of the infall. The names of codes with kinetic feedback are labeled in blue, the asterisk sign (*) denotes codes with delayed cooling and/or radiation pressure feedback, and the dagger sign ($\dagger$) denotes codes with superbubble feedback. The gas properties are computed within the primary halo's convex hull. On the bottom row, the solid lines represent the gas mass to total mass fraction ($M_\text{gas-total}/M_\text{all}$, left y-axis), and the dashed lines represent the star-forming gas mass to the total gas mass fraction ($M_\text{gas-sf}/M_\text{gas-total}$, right y-axis). The background colours represent the merger stages as indicated in Table~\ref{tab:merger_stages}. The stellar feedback scheme adopted in each code influences the star formation history during the interaction.}
    \label{fig:mass_sfr}
\end{figure*}

First, we investigate how star formation activity evolves through each dynamical phase of the interaction. Fig.~\ref{fig:mass_sfr} shows the evolution of stellar mass, SFR, gas mass, and gas mass fraction during the~1 Gyr following the start of each merger. Prior to the DM halo merger, only stars more bound to the primary than the secondary are used to compute the stellar mass and SFR; after the halo merger, all stars bound to the primary galaxy are used. The SFR is averaged over a 10-Myr timescale. The background colours of the plots represent the four merger stages defined in Table~\ref{tab:merger_stages}. The gas mass (fourth row) and gas mass fraction (fifth row) are computed within the primary halo's convex hull. Along with the total gas mass, we also show the amount of star-forming gas, defined as gas exceeding the \texttt{CosmoRun}'s gas density threshold for star formation ($n_\text{H} > n_\text{H,thres} = 1\,\text{cm}^{-3}$, or $\rho_\text{gas} > 2.2\times10^{-24} \text{g}\,\text{cm}^{-3}$). It is important to note that we do not impose a temperature limit on the star formation condition in \texttt{CosmoRun}. 

By construction, the codes have been calibrated to have approximately similar stellar mass (between $10^{9}$ and $5\times10^{9} \msun$) at $z = 4$ (Paper~\citetalias{Roca-Fabrega+2021}), which is near the coalescence timestep for most of the codes. Indeed, the stellar mass difference among the codes narrows throughout the interaction, decreasing from about~1 dex at $t_\text{start}$ to about 0.5 dex at $t_\text{start} + 1\,\text{Gyr}$. Nonetheless, the pattern of the stellar mass evolution exhibits noticeable discrepancies. For example, GIZMO demonstrates an abrupt surge in stellar mass while GEAR shows a steadier growth. This behaviour can be quantified with the star formation history (second row). We notice three groups of SFR patterns in the nine participating codes:
\begin{itemize}
    \item ART-I, GADGET-3, GADGET-4, and GIZMO: On the overall merger timescale, the SFR increases from the beginning of the first passage stage until about the first half of the coalescence stage, then decreases quickly in the second half and in the post-coalescence stage. In other words, the system experiences a pronounced, extended starburst. These four codes employ kinetic feedback in their feedback models (in combination with thermal feedback, not exclusively). The names of these codes in the subsequent figures are labeled in blue.
    \item ENZO, RAMSES, CHANGA, GEAR, and AREPO-T: On the overall merger timescale, the SFR increases from the beginning of the infall and into the post-coalescence stage with no significant drops. Small SFR enhancements still happen during the first passage stage; however, they are considerably smaller than the overall enhancement. These five codes use thermal feedback but not kinetic feedback in their prescription. The names of these codes in the subsequent figures are labeled in black. 
    \item ART-I, RAMSES, GADGET-3, and GEAR: On a short timescale between consecutive timesteps ($5-10$ Myr), the SFR is more bursty with small fluctuations. These codes use radiation pressure and/or delayed cooling in their stellar feedback models. The names of these codes in the subsequent figures are marked with an asterisk sign. 
\end{itemize}

It is important to emphasise that the short-timescale fluctuating SFR evolution of ART-I, RASMES, GADGET-3, and GEAR is independent of the merger-timescale SFR trend (increasing continuously versus peaking after the first apoapsis and declining). The reason is that radiation pressure and delayed cooling feedback supply excess pressure support that temporarily resists collapse on short timescales ($\leq 10$ Myr). The delayed cooling scheme (in RAMSES, GADGET-3, and GEAR) does this thermally, suspending radiative losses so that warm–hot gas builds up even at high density, while ART-I's radiation-pressure scheme adds a non-thermal pressure term to the surrounding dense gas during the first~5 Myr of a star particle \citep{Ceverino+2014}. These schemes lead to an accumulation of gas in a near star-forming state on a short timescale (compared to the SFR-averaged time of 10 Myr). Once the feedback fades, as cooling resumes or the ionizing sources age out, the gas collapses rapidly and continues to form stars. The resulting feedback re-pressurizes the surrounding gas, and the cycle makes gas repeatedly cross the star-formation threshold. Hence, this produces episodic bursts rather than steady star formation, as shown in the second row of Fig.~\ref{fig:mass_sfr}. As a result, even smooth gas inflow can translate into highly bursty star formation histories. We note that this short-timescale SFR burstiness persists till the end of the simulation, and the effect is less pronounced in GADGET-3 because it has a much shorter delayed cooling time compared to GEAR and RAMSES \citep[more details in Section~\ref{subsect:delayedcooling_gasproperties}]{Shimizu+2019}.

The correlation between the pattern of SFR evolution during the \textit{target} merger and the type of stellar feedback prescription, which persists across multiple code architectures in \texttt{CosmoRun}, strongly emphasises the importance of feedback implementation on merger predictions. Furthermore, we find no other systematic differences in the physical variables of the interaction (e.g., mass ratio, orbital trajectory, initial gas mass, shown in Table~\ref{tab:initial_state}) that can help explain this pattern.

\subsection{Gas properties during the merger}
\label{subsect:gas_properties_analysis}

\begin{figure*}[tbh]
    \centering
    \includegraphics[width=\linewidth]{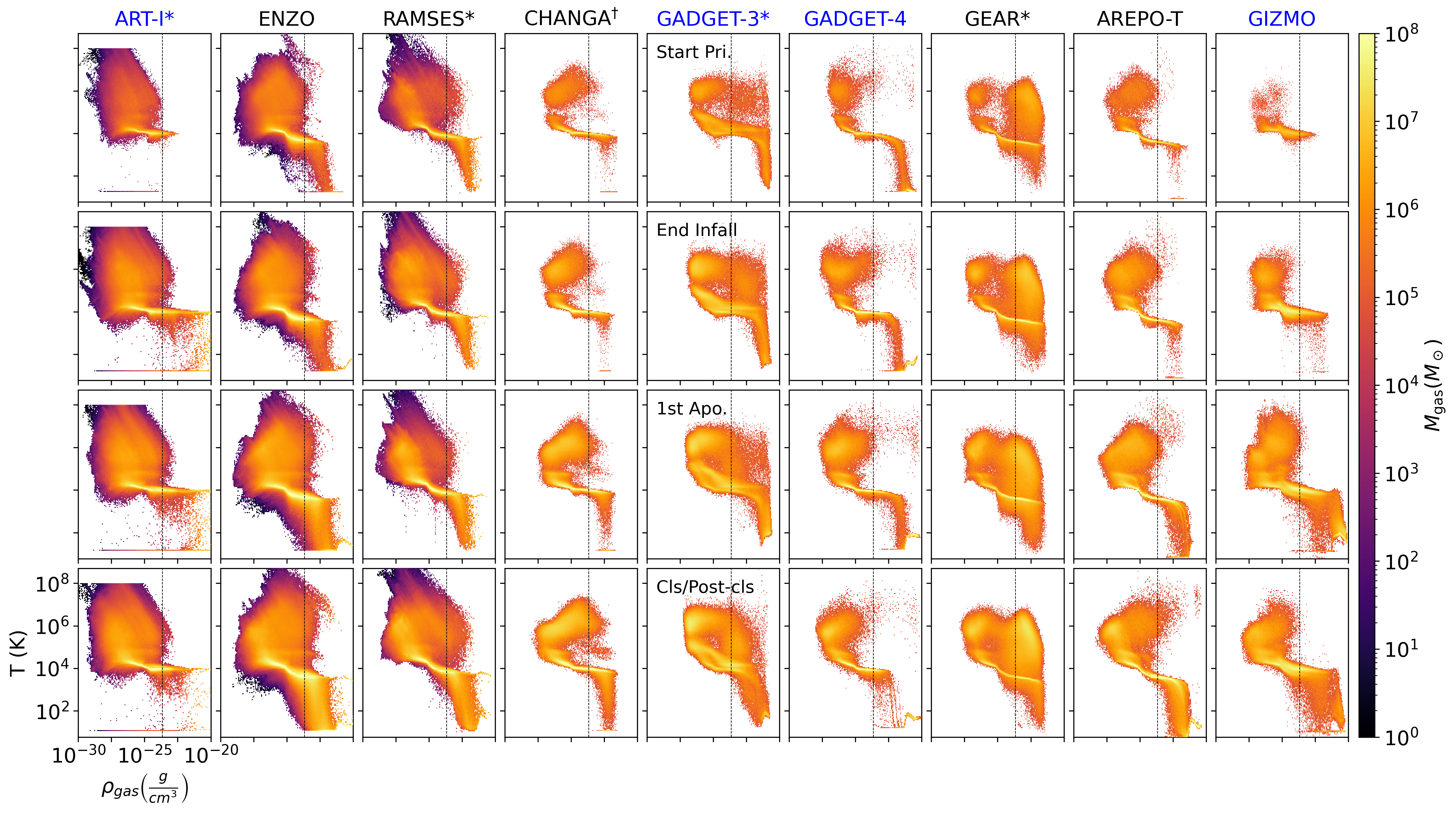}
    \caption{The gas phase plot showing the density-temperature distribution of all gas elements within the primary halo's convex hull at different times in the merging interaction. From top to bottom, we show the phase plots at $t_\text{start}$, at the end of the infall stage, at the end of the first passage stage ($t_\text{max}$), and at the timestep~0.6 Gyr after the first passage ("Cls/Post-cls"). For all codes but GEAR, this timestep lies in the post-coalescence stage, whereas it is in the coalescence stage for GEAR. After the infall stage, the primary halo's convex hull encompasses the interacting system. The vertical dotted lines represent the \texttt{CosmoRun}'s gas density threshold for star formation. The formatting of the code names follows that of Fig.~\ref{fig:mass_sfr}. The phase plots reflect the SFR patterns identified in Subsection~\ref{subsect:sfr_evolution_pattern}: gas in the codes using kinetic feedback (blue-coloured names) becomes densest and coldest around the first apoapsis, whereas the amount of cold dense gas in the codes using only thermal feedback (black-coloured names) increases more gradually through the merging interaction.}
    \label{fig:gas_phase_plot}
\end{figure*}

\begin{figure*}[tbh]
    \centering
    \includegraphics[width=\linewidth]{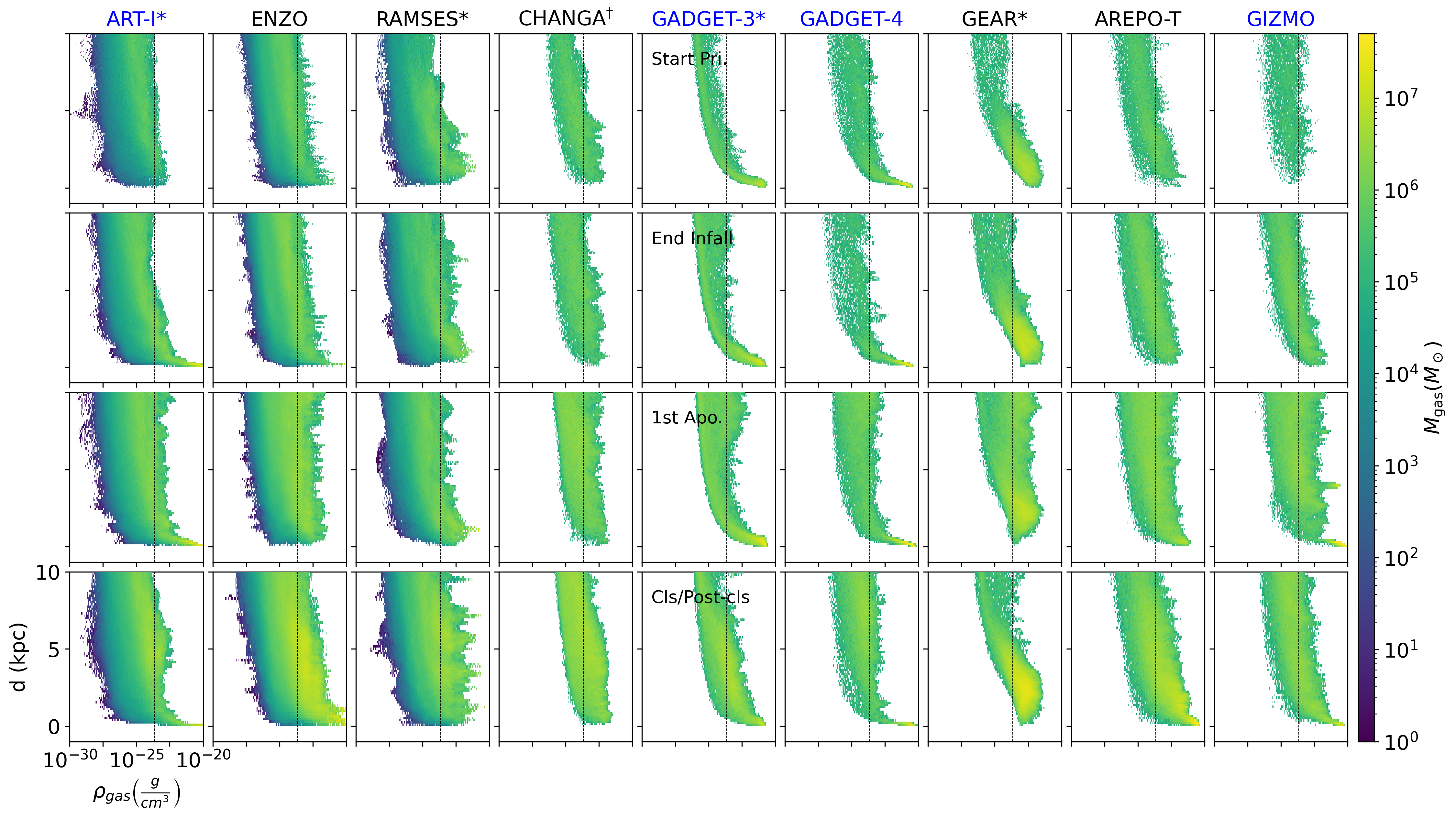}
    \caption{Similar to Fig.~\ref{fig:gas_phase_plot} but showing the distribution of density and distance to the galactic centre of the gas elements. Gas in codes with kinetic feedback (blue labels) reaches a high central density early in the interaction, whereas gas in only thermal-feedback codes reaches such density only near or after coalescence. Codes that adopt a delayed cooling or a superbubble scheme without kinetic feedback maintain a more uniform gas distribution throughout the merger.}
    \label{fig:gas_density_distance_plot}
\end{figure*}

To interpret the connection between the SFR pattern and the stellar feedback type, we examine the gas properties of the galaxies throughout the merging interaction. The third and fourth rows of Fig.~\ref{fig:mass_sfr} show the time evolution of the gas mass and gas mass fraction within the convex hull boundary of the primary halo. We do not find a clear correlation between the initial amount of gas in the system and the pattern or the intensity of the SFR. Indeed, even though the total gas mass agrees within~0.5 dex across all codes throughout the whole interaction, the SFR value range still spans about~$1\text{--}1.5 $ dex at any given timestep. Throughout the interaction, the system's total gas reservoir is continuously replenished by accretion, as evidenced by a gradual increase in the total gas mass. Even in ART-I, GADGET-4, and GIZMO, where a starburst episode leads to a decline in total gas mass, ongoing accretion subsequently restores the gas content after coalescence. Conversely, the amount of star-forming gas ($n_{H} > 1\,\text{cm}^{-3}$, or $\rho_\text{gas} > 2.2\times10^{-24} \text{g}\,\text{cm}^{-3}$) does not exhibit such quick restoration. Instead, the amount of star-forming gas generally traces the SFR pattern. To provide a more in-depth view of the gas, Figs.~\ref{fig:gas_phase_plot} and \ref{fig:gas_density_distance_plot} show the density-temperature and the density-radial distance distributions for all gas elements within the system. Sections~\ref{subsect:kinetic_gasproperties}, \ref{subsect:onlythermal_gasproperties}, and ~\ref{subsect:delayedcooling_gasproperties} show the gas properties of code groups exhibiting similar behaviour, and we discuss the physical explanation behind those properties in greater detail in Sections~\ref{subsect:SPH_trace_gas} and \ref{subsect:summary_gasproperties}. \footnote{For further interest, readers can refer to Paper~\citetalias{Strawn+2024} for an analysis of different gas phases in the galaxy and in the CGM over the full duration of each \texttt{CosmoRun} simulation.}

\subsubsection{Codes with kinetic feedback (ART-I, GADGET-3, GADGET-4, GIZMO)}
\label{subsect:kinetic_gasproperties}

For codes implemented with kinetic feedback (labeled in blue), the amount of star-forming gas peaks around the end of the first passage stage (i.e., at $t_\text{max}$), implying that the first periapsis drives a substantial amount of gas to high densities. The phase plot (Fig.~\ref{fig:gas_phase_plot}) and the density-distance plot (Fig.~\ref{fig:gas_density_distance_plot}) confirm this by showing more gas becoming denser and cooler, starting from the beginning of the interaction (first row), then the end of the infall stage (second row), and then the end of the first passage stage (third row). Indeed, by the end of the first passage stage, a larger amount of gas in ART-I, GADGET-3, GADGET-4, and GIZMO reaches a higher density and lower temperature than gas in codes that do not employ kinetic feedback. Fig.~\ref{fig:gas_density_distance_plot} also shows that the extremely dense gas clouds ($\rho_\text{gas} \approx 10^{-20} \text{g}\,\text{cm}^{-3}$) are located very close to the galactic centre. Hence, kinetic feedback allows gas to reach a higher density at the galactic centre more quickly during the merger. Slightly after the first passage stage, despite the increase in the total amount of gas, the amount of star-forming gas still decreases. The SFR of the system declines, as most of the cold dense gas is used during the starburst. This reduction is also reflected in the phase plots (Fig.~\ref{fig:gas_phase_plot}), where we see fewer dense cold gas elements in the post-coalescence timestep (fourth row) compared to the first apoapsis timestep (third row). The starburst also plays a role in halting star formation, as its intense feedback heats the interstellar medium (ISM) and prevents gas from further cooling and collapsing into the centre. Indeed, Fig.~\ref{fig:mass_sfr} (fourth row) shows that the ratio of star-forming gas to total gas decreases in the coalescence/post-coalescence stage, and the phase plot (Fig.~\ref{fig:gas_phase_plot}) displays a higher content of hot gas. 

\subsubsection{Codes with only thermal feedback (ENZO, AREPO-T)}
\label{subsect:onlythermal_gasproperties}

Unlike codes with kinetic feedback, codes implemented exclusively with thermal feedback show an increase in star-forming gas even after the two galaxies coalesce, as shown in the third row of Fig.~\ref{fig:mass_sfr}. This suggests a more gradual transition of gas into the star-forming state compared to codes with kinetic feedback. Indeed, the phase plots (Fig.~\ref{fig:gas_phase_plot}) of ENZO and AREPO-T show that gas continues to become denser and colder even after the two galaxies coalesce. The fourth row of Fig.~\ref{fig:mass_sfr} displays that the ratio of star-forming gas to total gas mass (dashed lines, right y-axis) also gets considerably larger over time, proving that this increase in star-forming gas is not just due to the increase in the total amount of gas in the system (due to accretion), but due to more gas collapsing into the galactic centre and becoming dense and cold. Fig.~\ref{fig:gas_density_distance_plot} verifies this by showing a larger amount of gas at the inner radius over time. 

\subsubsection{Codes with delayed cooling or superbubble scheme, and without kinetic feedback (RAMSES, CHANGA, GEAR)}
\label{subsect:delayedcooling_gasproperties}

As discussed in the second bullet point in Section~\ref{subsect:sfr_evolution_pattern}, on the overall merger timescale, RAMSES, CHANGA, and GEAR still show an increasing trend of SFR into the post-coalescence stage, similar to ENZO and AREPO-T. The implementation of the delayed cooling scheme (for RAMSES and GEAR) and the superbubble scheme (for CHANGA), however, introduces an additional effect. 

For RAMSES and GEAR, we do not see the star-forming gas mass increase significantly throughout the interaction, compared to codes with only thermal feedback (third row of Fig.~\ref{fig:mass_sfr}). The fourth row of Fig.~\ref{fig:mass_sfr} also shows that the fraction of star-forming gas to total gas mass slightly decreases throughout the merging interaction in both simulations. The delayed cooling effect is clearly visible in the phase plots (Fig.~\ref{fig:gas_phase_plot}), where we see the accumulation of warm–hot gas in a relatively dense state. Yet, the delayed cooling scheme limits the formation of very dense gas with $\rho_\text{gas} > 10^{-21} \text{g}\,\text{cm}^{-3}$, which is present in the other six codes during the examined time period (noticeably seen in Fig.~\ref{fig:gas_density_distance_plot}). With a longer constant delay time of~$5-10\,\text{Myr}$, the delayed cooling scheme in RAMSES and GEAR pressurizes the dense gas by keeping it hot and out of equilibrium. Because cooling is turned off, the cooling time becomes infinite, leading to gas maintaining its thermal motions and lengthening the collapse time. Consequently, this suppression of cooling inhibits the concentration of cold and very dense gas, thus mitigating strong and prolonged starbursts that are produced in other simulations. Instead, a more spatially extended gas distribution is created, as displayed by Fig.~\ref{fig:gas_density_distance_plot}. The same behaviour was indeed found in another simulation of high-redshift galaxies using the delayed cooling stellar feedback scheme underlying the \texttt{CosmoRun} RAMSES model \citep{Dubois+2015}. We emphasise that the suppression of strong, prolonged starbursts due to the delayed cooling scheme should be distinguished from the episodic bursts it induces, as mentioned in Section~\ref{subsect:sfr_evolution_pattern}. These episodic bursts occur on shorter timescales ($\approx$ 5-10 Myr) and exhibit smaller amplitudes ($\approx 2\text{--}7\,\msun/\text{yr}$) compared to the starbursts we see in codes with kinetic feedback ($\approx$ 500 Myr in timescale and $> 10\,\msun/\text{yr}$ in amplitude). 

It is important to note that GADGET-3 also employs a delayed cooling scheme, which is also reflected by the accumulation of hot, dense gas in its phase plots. But instead of using a fixed value for the delay time, GADGET-3 adjusts it based on the hot phase duration of supernova feedback, $t_\text{delayed} = t_\text{hot} \propto n_{0}^{0.27}P_{0}^{-0.64}$, where $n_{0}$ is the ambient hydrogen density and $P_{0}$ is the ambient gas pressure \citep{Shimizu+2019}. As pointed out in Paper~\citetalias{Jung+2025a}, this dependence of the delayed cooling time on density and pressure allows warm-hot dense gas in GADGET-3 to have $t_\text{delayed} \leq 0.1 \,\text{Myr}$. This short $t_\text{delayed}$ allows gas particles to cool and form cold, dense gas more effectively than in RAMSES or GEAR, whose $t_\text{delay}$ is fixed at $5-10$ Myr. Thus, GADGET-3 more closely traces the behaviour of codes implementing kinetic feedback rather than those adopting a delayed cooling feedback. 

Similar to RAMSES and GEAR, the superbubble prescription of CHANGA prevents gas from reaching a very high density ($\geq 10^{-21} \text{g}\,\text{cm}^{-3}$) throughout the whole merging process. Instead of turning off gas cooling, the superbubble scheme causes very strong feedback-driven outflows that expel gas into the circumgalactic medium (CGM), preventing gas from falling into the galactic centre (see fig.~12 of Paper~\citetalias{Jung+2025a}). This outflow eventually cools down and re-accretes back onto the galaxy, creating a delayed starburst about 0.5 Gyr after coalescence. We discuss this delayed starburst further in Appendix~\ref{subsect:CHANGA_delayed_starburst}.

Lastly, it is important to note that our results on the SFR patterns and gas properties may be most relevant to a major merger with an approximate mass range and orbital configuration as the \textit{target} merger. Because \texttt{CosmoRun} is a zoom-in simulation suite primarily focusing on one galaxy, more studies need to be done to investigate whether this connection between stellar feedback models and SFR evolution pattern exists in other galaxy merger configurations. 

\subsubsection{Gas particle tracking}
\label{subsect:SPH_trace_gas}

\begin{figure*}[tbh]
    \centering
    \includegraphics[width=\linewidth]{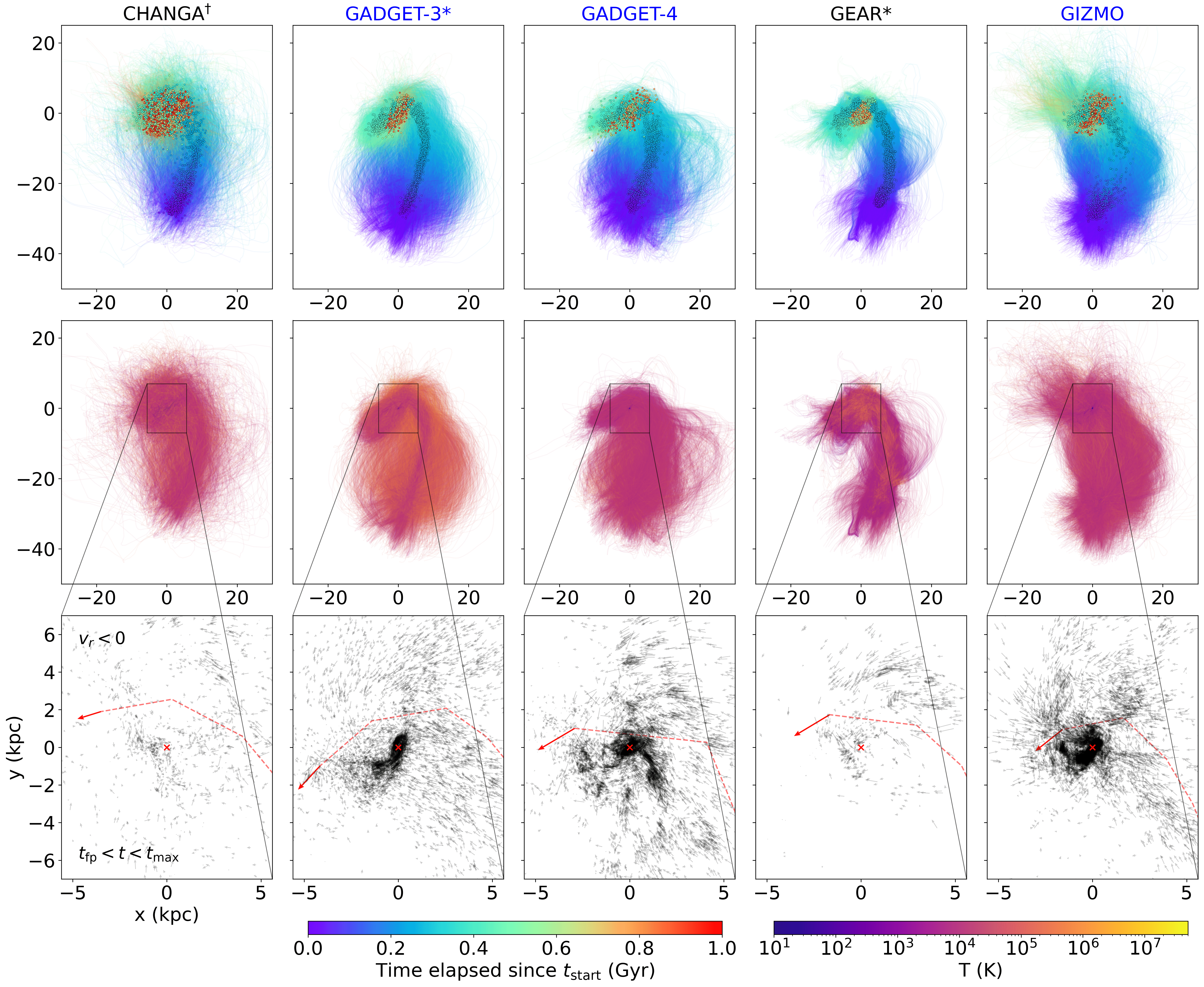}
	\caption{Tracking the gas particles from the pre-infall secondary halo that turn into star particles within $\approx 1\,\text{Gyr}$ after $t_\text{start}$. The plot origin (red cross) is at the primary galaxy's centre. Top row: the trajectories of each gas particle, coloured by elapsed time since $t_\text{start}$. The scatter points show where gas particles turn into star particles, with colours denoting the star formation time relative to $t_\text{start}$. Middle row: the trajectories coloured by the gas temperature. Bottom row: a zoomed-in view of the primary galaxy's centre at a timestep between the first periapsis ($t_\text{fp}$) and the first apoapsis ($t_\text{max}$), displaying the velocity vectors of gas particles with negative radial velocity components. In each panel, the red dashed line represents the merger's orbital path, and the red arrow shows the velocity vector of the secondary galaxy's stellar core. Codes with kinetic feedback (GADGET-3, GADGET-4, and GIZMO) have gas accrete from the secondary galaxy onto the primary galaxy more effectively and trigger earlier starbursts. In contrast, the other two codes suppress gas cooling through the delayed cooling (GEAR) and the superbubble (CHANGA) scheme, resulting in higher random thermal motions that hinder gas accretion to the galactic centre.}
\label{fig:track_gas_particles_turning_into_stars_fromsecondaryHalo}
\end{figure*}

\begin{figure}
    \centering
    \includegraphics[width=0.95\linewidth]{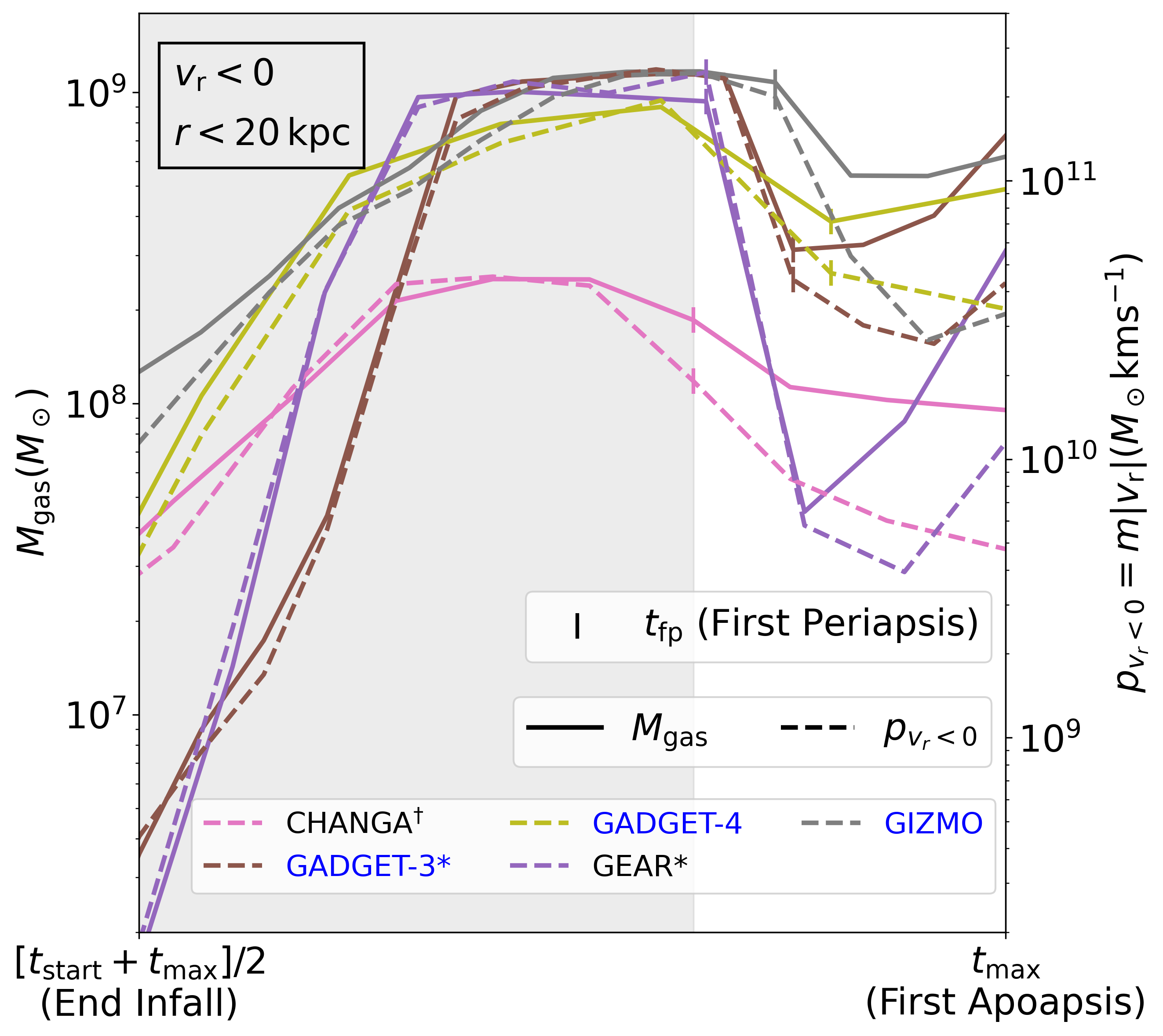}
    \caption{The time evolution of total mass (solid lines) and total radial momentum (dashed lines) of pre-star gas particles from the secondary galaxy with $v_r < 0$. We focus this plot on the first passage stage and on gas within a 20-kpc radius from the primary galaxy's centre, as the starbursts in codes with kinetic feedback begin in this stage and within that radius (first row of Fig.~\ref{fig:track_gas_particles_turning_into_stars_fromsecondaryHalo}). In each code, the short vertical line denotes the first periapsis timestep. The non-shaded region highlights the approximate interval between $t_\text{fp}$ and $t_\text{max}$ (since this interval slightly differs among codes, the shaded boundary is set to the first periapsis by CHANGA). Between the first periapsis and first apoapsis, kinetic feedback codes (GADGET-3, GADGET-4, and GIZMO) transfer substantially greater gas mass and momentum from the secondary to the primary galaxy than the other two codes, despite the galaxies receding from one another during this interval. This matches our observation in Fig.~\ref{fig:track_gas_particles_turning_into_stars_fromsecondaryHalo}.}
    \label{fig:infalling_momentum}
\end{figure}

In Section~\ref{subsect:sfr_evolution_pattern}, we identify two different SFR evolution patterns on the overall merger timescale during the \textit{target} merger across the codes. These differences can be attributed to the feedback prescriptions: adding kinetic feedback can drive gas to higher density more quickly and cause an earlier starburst, whereas using purely thermal feedback heats the gas, resulting in a continuously increasing SFR as gas takes a longer time to cool. To further examine this explanation, for the particle-based codes, we traced the trajectories of the gas particles that turn into star particles during the merger event. This tracking can help us study how gas accretes and collapses into the galactic centre to form stars. In the \texttt{CosmoRun} suite, we have four particle-based codes: CHANGA, GADGET-3, GADGET-4, and GEAR. In addition, the hybrid, meshless code GIZMO also employs a Lagrangian approach for its fluid elements, and each element's position can be traced through time. The \texttt{CosmoRun}'s common star formation prescription requires that when a new star particle spawns in particle-based codes, it will inherit the mass of the parent gas particle. Moreover, except for CHANGA, the particle ID of a newly born star particle will be inherited or computable from the ID of its parent gas particle \footnote{In GADGET-4, a gas particle can form multiple star particles and does not necessarily disappear after forming one. The ID of each star particle can be traced back to its unique parent gas particle using a simple equation. In Fig.~\ref{fig:track_gas_particles_turning_into_stars_fromsecondaryHalo}, we only plot the gas trajectories up to a Gyr after $t_\text{start}$ or until they disappear.}, allowing us to pinpoint all the gas elements that contribute to star formation. For CHANGA, we cannot trace the stellar IDs back to the gas IDs, so new stars are identified by gas particles that disappear during the merger. 

Because the CHANGA simulation in \texttt{CosmoRun} employs an identical initial stellar mass for all newborn stars ($\approx 5.65\times10^{4} \msun = m_\text{gas, IC}$), one gas particle with enough mass can still exist after forming one star particle, as long as its remaining mass is larger than $m_\text{gas, IC}$. Therefore, tracing only the disappearing gas particles may not capture the full population of gas contributing to the merger-induced starburst. Nevertheless, because our goal is not to precisely quantify the conversion of gas mass into stellar mass but rather to provide a qualitative interpretation, and because a majority of gas particles disappear after forming stars ($\approx$~80\%), back-tracing only the disappearing gas particles in CHANGA is sufficient to elucidate the gas dynamics that lead to star formation during the merger. Another limitation that prevents us from quantifying the fractional contribution of each gas source (main galaxy versus secondary galaxy versus accretion) is diffusion. In GAGDET-3 and GEAR, diffusion is implicit in the smoothing procedure, meaning that there is no exchange of mass between gas particles. Alternatively, CHANGA and GADGET-4 explicitly model metal diffusion (even though GADGET-4 has a very weak diffusion coefficient of~0.002, which has minimal impact on mass exchange), while GIZMO models diffusion self-consistently using unstructured cells. Diffusion processes allow thermodynamic quantities to be redistributed among neighboring gas particles. This mixing effect is expected to be most pronounced in dense gas regions. Consequently, the back-tracked particle histories should be interpreted as qualitative tracers of the gas dynamics leading to star formation, rather than as exact thermodynamic and mass histories of the fluid elements.

Fig.~\ref{fig:track_gas_particles_turning_into_stars_fromsecondaryHalo} displays the gas trajectories coloured by time (top row) and temperature (middle row) of the gas particles from the secondary galaxy that turn into star particles within $\approx$~1 Gyr of $t_\text{start}$ (matching the interval covered from the first to fourth rows of Figs.~\ref{fig:gas_phase_plot} and ~\ref{fig:gas_density_distance_plot}). The gas particles are selected from within the secondary halo's convex hull at $t_\text{pre-infall}$, and we do not plot gas particles that do not form stars. For brevity, we will refer to the gas particles that spawn stars within $\approx$~1 Gyr of $t_\text{start}$ as "pre-star gas". The top row of the figure also displays the locations of the star particles when they first form from the pre-star gas particles. The scatter points are coloured by the stars' formation time, with violet indicating stars formed at $t_\text{start}$ and bright red indicating stars formed around $\approx t_\text{start} + 1\text{Gyr}$. We can notice that the secondary galaxy continuously forms stars during the interaction. We also verified that there is no significant difference in the size of secondary galaxies among the codes.

As seen in the second row of Fig.~\ref{fig:mass_sfr}, the strong, prolonged starburst events in codes with kinetic feedback (blue-named codes) begin in the first passage stage of the interaction (yellow background). By examining the motions of pre-star gas particles in this stage, we can better examine the origin of these starburst events. In the bottom row of Fig.~\ref{fig:track_gas_particles_turning_into_stars_fromsecondaryHalo}, we zoom in on the primary galaxy's centre and display the locations as well as the velocity vectors of the pre-star gas particles with a negative radial velocity component with respect to the primary galaxy ($v_r$). A negative $v_r$ implies an infalling motion or accretion to the primary galaxy, and a system with more accretion can drive an increase in the local gas density and SFR more efficiently. We also chose a timestep between the first periapsis ($t_\text{fp}$) and the first apoapsis ($t_\text{max}$) to disentangle the gas accretion during the first passage stage from the large-scale infall driven by the interaction's direction. We observe a stark difference between codes with kinetic feedback (GADGET-3, GADGET-4, and GIZMO) and codes without kinetic feedback (CHANGA and GEAR). After the first periapsis, there are considerably more gas particles with $v_{r} < 0$ from the secondary galaxy in GADGET-3, GADGET-4, and GIZMO. These gas particles move distinctly differently from their galaxy's bulk motion. At the chosen timestep, even though the secondary galaxy is receding from the primary galaxy (the velocity vector of the secondary galaxy's stellar core is shown by red arrows), the plotted gas particles still flow into the primary galaxy. The accretion from the secondary galaxy to the primary galaxy is thus very noticeable. In addition, the dense region in the panels of GADGET-3, GADGET-4, and GIZMO suggests that, after accretion, gas from the secondary galaxy continues to collapse further into the primary galaxy's centre. 

We have two hypotheses for why kinetic feedback enhances gas accretion from the secondary galaxy onto the primary. First, \citet{Chaikin+2023} find that at a fixed total feedback energy, increasing the relative contribution of the kinetic feedback channel helps retain more gas in the ISM and limit galactic-scale outflows. They argue that, by channeling part of the supernova energy into kinetic injection, the model curtails the growth of hot superbubbles. This suppression then inhibits the development of a hot ISM phase and prevents gas from being ejected from the galaxy. As a result, in our case, gas can accrete onto the main galaxy more efficiently during the first passage. The second hypothesis is that, for codes with kinetic feedback, gas elements surrounding an SN event receive a momentum change. While a purely thermal feedback scheme injects energy isotropically and raises the internal energy of neighboring gas elements, kinetic feedback imparts a momentum kick to the surrounding gas with a direction away from the SN. When the momentum kick is directed opposite to a gas particle's velocity (for example, a supernova in the primary galaxy exerts a momentum kick in an opposite direction to gas outflow from the secondary galaxy during the first passage), the net kinetic energy of that gas particle is reduced, effectively removing its orbital energy. As a result, the gas particle can fall into the primary galaxy's centre more easily. Dynamical friction can also play a role in facilitating this accretion. 
With more gas being accreted, gas can reach a higher central density, and a strong starburst can happen. This feature is less apparent in CHANGA and GEAR because of their strong heating schemes, causing gas to expand and not collapse into the galactic centre.

In contrast, thermal feedback, combined with the superbubble scheme in CHANGA and delayed cooling in GEAR, suppresses gas cooling during the merger timescale. We can observe this heating effect in the middle row of Fig.~\ref{fig:track_gas_particles_turning_into_stars_fromsecondaryHalo}. Unlike GADGET-3, GADGET-4, and GIZMO, we do not observe a cold ($< 100K$), dense, point-like region into which the pre-star gas collapses in CHANGA and GEAR. Instead, the very central regions in these two codes are permeated by hot and warm gas. The effect is particularly pronounced in GEAR, where the centre of the secondary galaxy is substantially warmer than its outer regions because of the delayed cooling scheme. This suppression allows thermal energy to effectively cancel dynamical friction, resulting in a domination of thermal random motion over inflow. Consequently, gas particles are inhibited from funneling toward and concentrating in the galactic nucleus, making it less likely for a strong starburst to happen during the pericentric passage. In the GEAR and CHANGA panels in the bottom row of Fig.~\ref{fig:track_gas_particles_turning_into_stars_fromsecondaryHalo}, the amount of gas with a negative radial velocity component is significantly lower than their kinetic-feedback counterparts. We further demonstrate this in Fig.~\ref{fig:infalling_momentum}, where we plot the total mass (solid lines) and total radial momentum component (dashed lines) of gas particles with negative radial velocity. As most stars form within~20 kpc from the primary galaxy's centre, we imposed a radius limit to avoid contamination of gas particles too far away in the plot. Fig.~\ref{fig:infalling_momentum} shows that between $t_\text{fp}$ and $t_\text{max}$, both the total mass and the total momentum of gas with $v_{r} < 0$ are about 4-10 times smaller for codes without kinetic feedback. For CHANGA, this difference is still larger than our underestimation of its pre-star gas particles ($\approx$ 1.25 times), as discussed earlier in this section. We note that before $t_\text{fp}$, the negative radial velocities are dominated by the bulk infall of the secondary galaxy. Therefore, this period on the plot does not reflect the stellar-feedback-driven gas inflow component.

Even though we argue that gas in GEAR has more random motion, the trajectories of GEAR's pre-star gas particles follow the orbital path more closely (top row of Fig.~\ref{fig:track_gas_particles_turning_into_stars_fromsecondaryHalo}). We want to emphasise that this behaviour arises because only gas particles that turned into stars during the merger are plotted. For GEAR, only gas in the central regions of the secondary galaxy remains sufficiently dense to later form stars during the merger. We have verified that, on average, the remaining gas particles that do not form stars attain higher temperatures, move more randomly, and get expelled to larger distances beyond the orbital path.  

\subsubsection{Summary of the connection between the stellar feedback model, gas properties, and the SFR pattern}
\label{subsect:summary_gasproperties}

After gathering the evidence from analysing the gas properties as shown in the previous subsections, we summarise our physical interpretation of why different stellar feedback models result in different SFR patterns during the merger as follows. In simulations that implement kinetic feedback (ART-I, GADGET-3, GADGET-4, and GIZMO), gas from the secondary galaxy can be injected with opposite-direction momentum to accrete more effectively onto the primary galaxy, especially during the first periapsis. This mechanism provides a quicker channel for cold gas to accrete and funnel into the centre during the interaction, leading to a more rapid increase in the SFR following infall. \cite{Chaikin+2023} also showed that when the fractional contribution of the kinetic feedback in the total feedback energy increases, more gas will stay in the ISM instead of being ejected outside the galaxy. In addition, GADGET-3 and GADGET-4 develop a disc after the pre-merger starburst event at $z = 5.5$ (see Appendix~\ref{appendix:GADGET-preMerger-Starburst}, Paper~\citetalias{Jung+2025a}, and Paper IX - Part 2). The presence of a disc makes the galaxy susceptible to gravitational instabilities \citep{Toomre+1964}, which can also play a role in triggering earlier star formation. By contrast, using thermal feedback injects energy only in the form of heat, producing hot diffuse outflows in the form of bubbles that can break out of the ISM into the CGM (refer to Paper~\citetalias{Strawn+2024} for a detailed comparison of the AGORA \texttt{CosmoRun} suite regarding the CGM). \cite{Keller+2020} pointed out that if the bubbles attain sufficiently high entropy, they experience an outward acceleration due to a local buoyant force. This force lifts the bubbles to large galactocentric radii, allowing them to remain in the CGM for extended periods, and therefore delaying the cooling and collapsing of gas into the galactic centre after the merger. Once the outflow gas cools down, it can gradually accrete back onto the galaxy and lead to a continuously increasing SFR, particularly as seen in ENZO and AREPO-T, codes that use only thermal feedback. Furthermore, using thermal feedback without kinetic feedback increases the random motion of gas elements, making it more challenging for the gas to flow from the secondary galaxy to the primary galaxy during the first periapsis, thus limiting the occurrence of a strong burst during this merger stage. Notably, ENZO and AREPO-T have the highest thermal energy within all codes ($5\times 10^{52}\,{\rm ergs/SN}$ and $2\times 10^{52}\,{\rm ergs/SN}$, respectively, see Table~\ref{tab:feedback}), which further amplifies the effect of thermal feedback. When the thermal feedback (without kinetic feedback) is coupled with the delayed cooling scheme (RAMSES and GEAR) or the superbubble scheme (CHANGA), these schemes further hinder the formation of very cold dense gas at the galactic centre, either by turning off cooling or strongly ejecting gas away from the galaxy. Hence, merger-driven inflow is suppressed, and stars formed during the interaction in these codes are not compactly distributed. Details on the effect of the \textit{target} merger on stellar morphology are discussed in Paper IX - Part 2. 

\subsection{Burst fraction}
\label{subsect:burst_fraction}

\begin{figure}[tbh]
    \centering
    \includegraphics[width=\columnwidth]{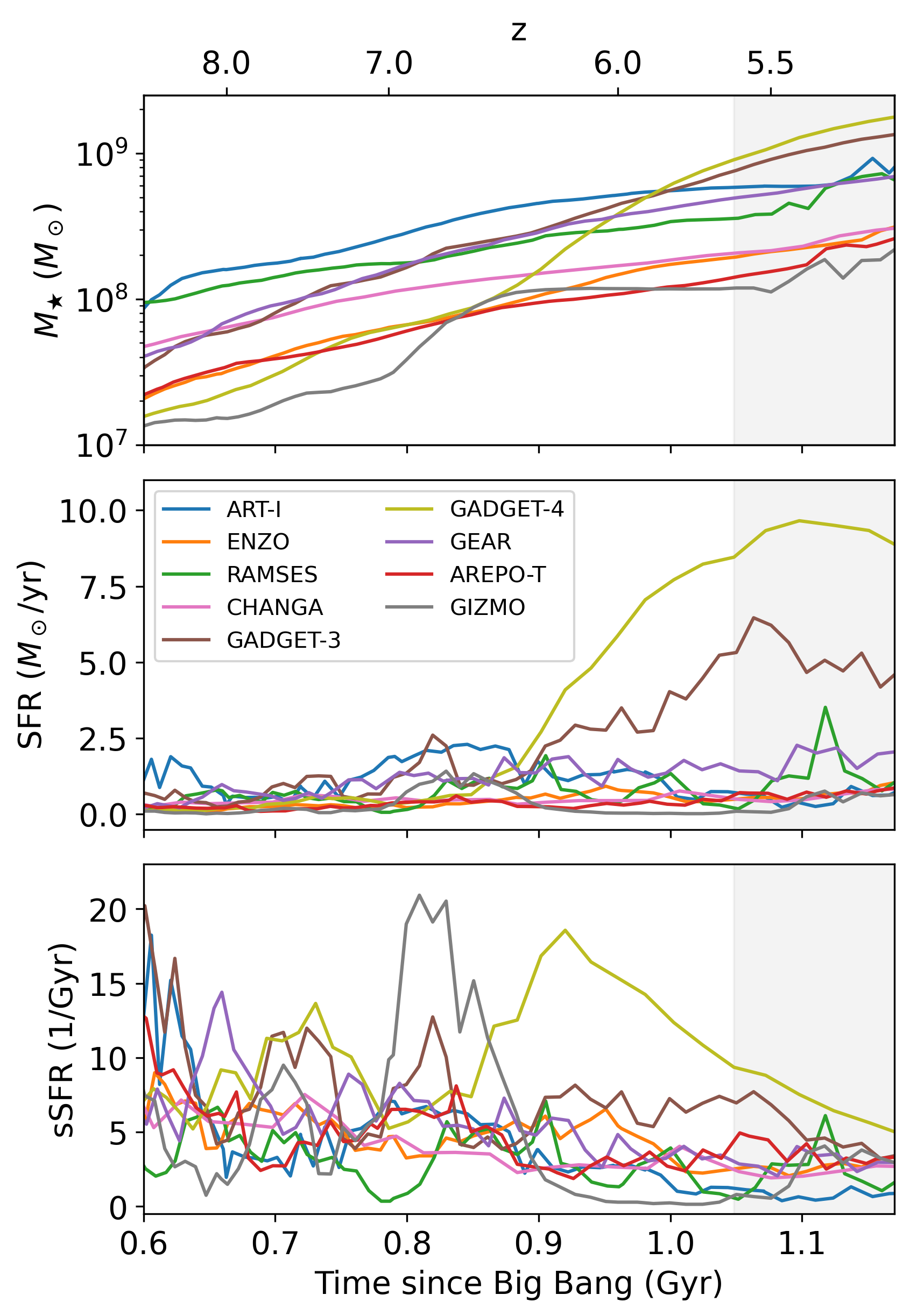}
    \caption{The stellar mass (top row), SFR (middle row), and sSFR (bottom row) of the main galaxy before and shortly after the infall of the \textit{target} merger. The vertical grey band represents the range of $t_\text{start}$ in all nine codes. Except for GADGET-3 and GADGET-4, the SFR and sSFR in the other codes remain relatively low and stable right before the \textit{target} merger, allowing us to reliably calculate the baseline sSFR and the burst fraction. GADGET-3 and GADGET-4 display strong starbursts shortly before the \textit{target} merger; therefore, we cannot robustly calculate their burst fraction.}
    \label{fig:mass_sfr_evolution_premerger}
\end{figure}

\begin{figure*}[tbh]
    \centering
    \includegraphics[width=\linewidth]{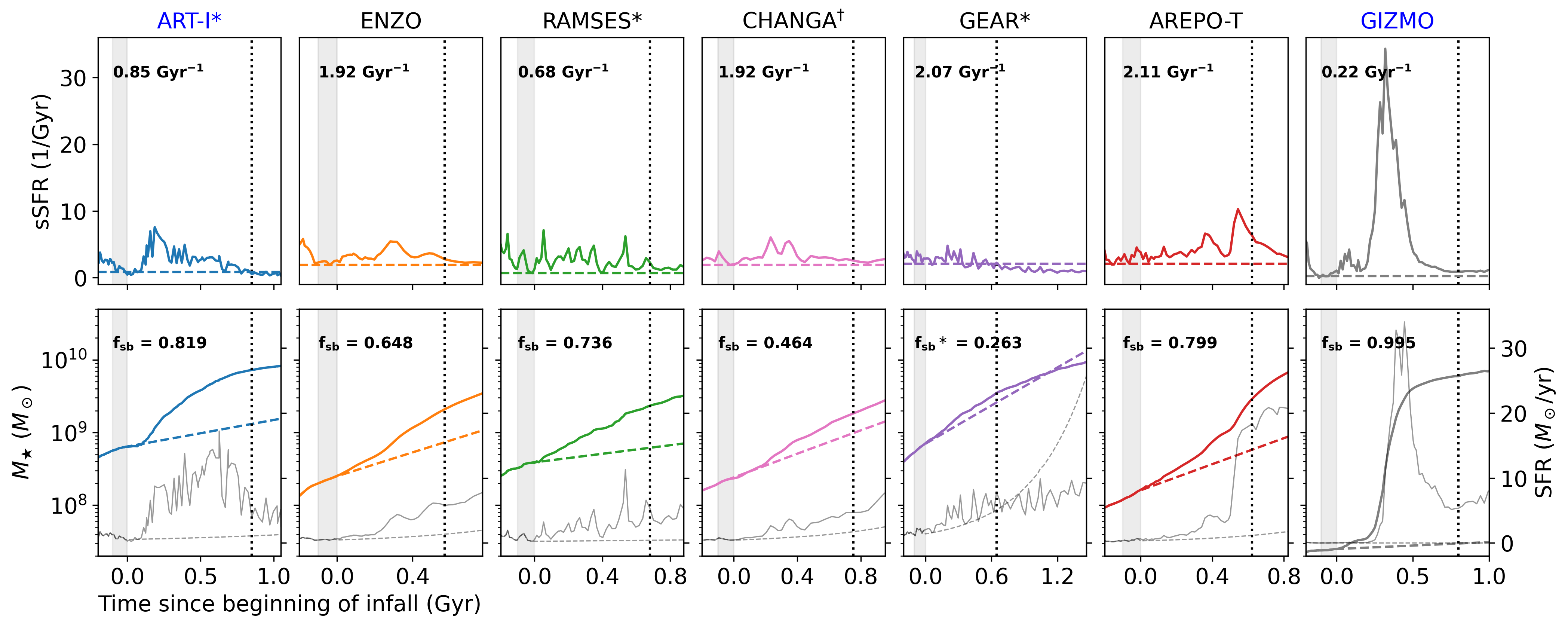}
    \caption{The sSFR (top row), stellar mass (bottom row, solid coloured line), and SFR (bottom row, gray solid line) of the primary galaxy, computed after excluding all stars that are at any time more bound to the secondary galaxy than the primary. The constant baseline sSFR (the sSFR assuming that the main galaxy evolves in isolation) is computed as the lowest sSFR over the 100-Myr period before the \textit{target} merger's infall (gray vertical band). The baseline sSFR values are shown by the dashed coloured lines and quoted in the top-row subplots. This baseline sSFR is used to calculate the baseline SFR (dashed gray lines) and the baseline stellar mass (dashed coloured lines) in the bottom row. Except GEAR, the burst fraction ($f_\text{sb}$) is evaluated at coalescence (vertical dotted line) and is about $0.4\text{--}0.8$. GADGET-3 and GADGET-4 are excluded from this computation due to having another strong starburst preceding the \textit{target} merger that prevents us from robustly calculating the baseline sSFR.}
    \label{fig:burst_fraction}
\end{figure*}

\begin{figure}[tbh]
    \centering
    \includegraphics[width=\linewidth]{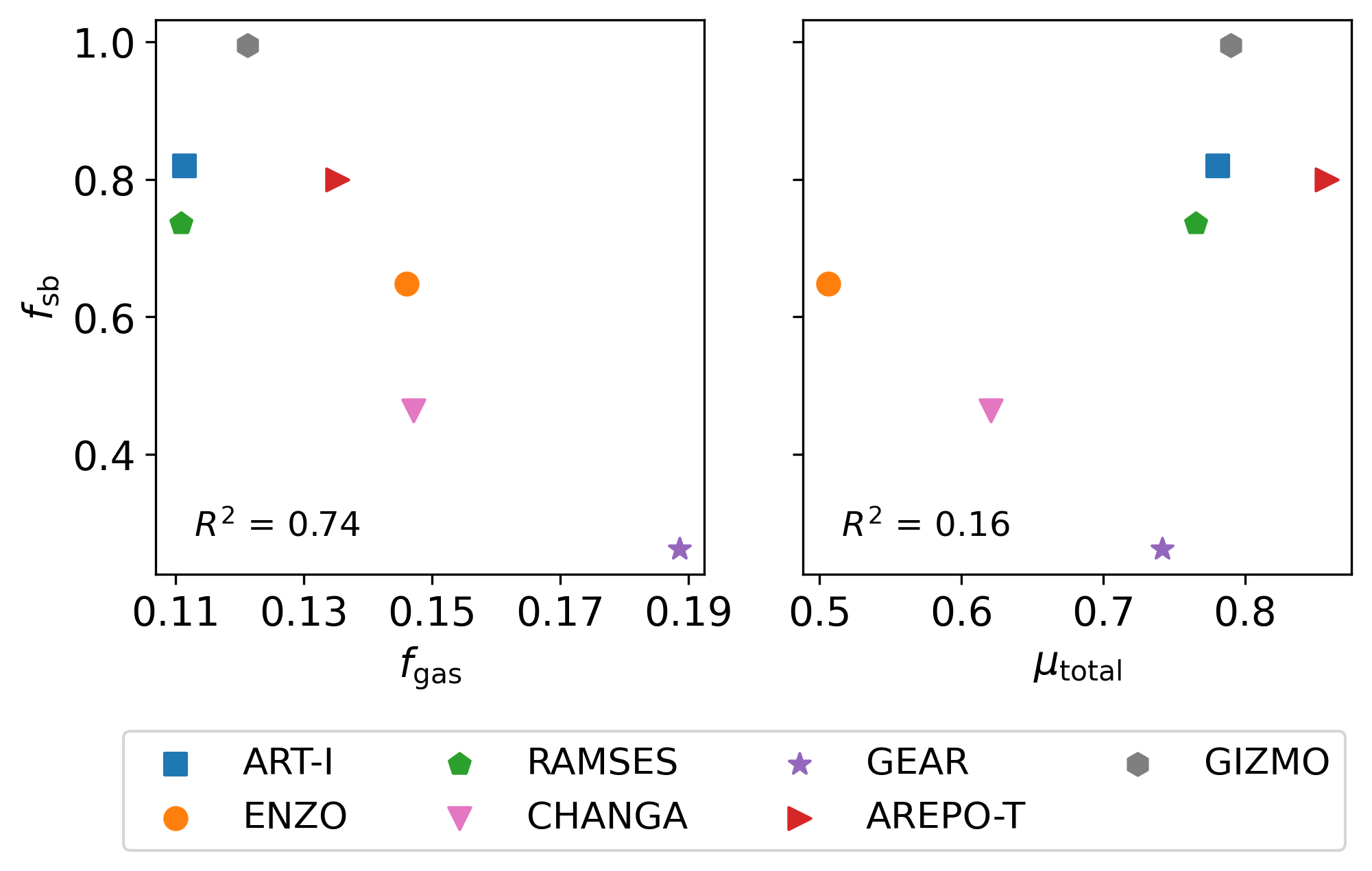}
    \caption{The relationship between the burst fractions ($f_\text{sb}$) and the system's gas fraction ($f_\text{gas}$, left) as well as the total mass ratio ($\mu_\text{total}$, right). The $R^{2}$ value of a simple linear regression is shown for each subplot. We find that the burst fraction is negatively correlated with the gas fraction and very weakly correlated with the total mass ratio. At sufficiently small $f_\text{gas}$, however, $f_\text{sb}$ may also be suppressed.}
    \label{fig:burst_fraction_correlation_with_PreMerger}
\end{figure}

The burst fraction (the fraction of stars that form during a merger but would not have formed if the galaxy were isolated, also called the burst efficiency) of a galaxy merger is typically defined relative to a control sample. The control sample can be the same galaxy evolved in isolation within the same time interval in the case of idealised simulations \citep{Cox+2008}, or a sample of isolated galaxies at the same stellar mass and local environment density for observational studies \citep{Reeves+2023a}. Using a control sample helps separate the enhancement in star formation directly attributable to the merging interaction from that driven by secular evolution and/or filamentary accretion. Nonetheless, such a setup is not feasible in fully cosmological, zoom-in simulations targeting a single galaxy. 

In this study, we propose a new way to estimate the burst fraction without relying on a control galaxy sample. We assume that in an environment with a relatively continuous and smooth gas accretion from the cosmic filaments and without any halo mergers with a mass ratio $> 1:10$ (below which the difference between a merger and accretion becomes increasingly blurred), the sSFR of a galaxy remains constant. This assumption is more valid for a shorter amount of time when the expansion of the universe does not significantly affect the accretion rate, and for a more massive galaxy whose stellar feedback less alters its mass and whose gas density is much larger than the filament's density. In our codes, the merger timescale is about $0.6\text{--}0.8\, \text{Gyr}$, and the pre-infall galaxy is relatively massive ($M_\bigstar \approx 10^{9} \msun$) with a high central density ($\rho \approx 10^{-23}\text{--}10^{-22} \text{g}\,\text{cm}^{-3}$). After selecting our constant "baseline" sSFR, we used that baseline sSFR to predict the baseline SFR and stellar mass evolution that the galaxy would have had if the \textit{target} merger did not occur. The selected baseline sSFR needs to be carefully chosen to be representative of the galaxy when it is free from other mergers. In addition, it should preferably be chosen shortly before the \textit{target} merger, so that the value best reflects the pre-merger mass of the galaxy. As shown in Fig.~\ref{fig:merger_tree}, the main halo does not experience any mergers with $\mu_\text{DM}$~>~1:10 in the last $\approx$~100 Myr before the infall of the \textit{target} merger. To further ensure that the galaxy's star formation activity is stabilized, we show in Fig.~\ref{fig:mass_sfr_evolution_premerger} the evolution of stellar mass, SFR, and sSFR before $t_\text{start}$ of each code. The vertical grey band marks the range of $t_\text{start}$ of all nine codes. Except for GADGET-3 and GADGET-4, the other seven \texttt{CosmoRun} codes show a relatively low and stable SFR and sSFR during~100~Myr before the \textit{target} merger. The stellar mass plot also does not display any sudden jumps, showing that there are no other considerable galaxy mergers that can violate our assumption when determining the baseline sSFR. It is important to note that the main galaxy in GIZMO undergoes an intense starburst at $t \approx 0.8$ Gyr after the Big Bang. Nevertheless, in the 100 Myr before the \textit{target} merger, the galaxy had already relaxed after this starburst, and hence, our baseline sSFR of GIZMO is not influenced by it. 

As mentioned above, GADGET-3 and GADGET-4 do not show a stable SFR right before the \textit{target} merger. At $t \approx 0.9$ Gyr after the Big Bang, the galaxy in both codes undergoes a strong starburst that continues considerably after the infall of the \textit{target} merger. Indeed, the burst in GADGET-4 is so strong that it increases the main galaxy's stellar mass by an order of magnitude, elevating the galaxy from one of the least massive galaxies among the codes to the most massive at $z = 6$. Moreover, this starburst results in a significantly more centrally concentrated gas distribution in the main galaxies of GADGET-3 and GADGET-4, accompanied by the formation of a galactic disc (fig.~4 of Paper~\citetalias{Jung+2025a}). Our hypothesis to explain the starburst is the collapse of gas to the galactic center due to the gas's low collapse time and/or an instability introduced by a very minor merger with a DM mass ratio of $\approx 0.03$. A detailed investigation of these starbursts in GADGET-3 and GADGET-4 is discussed in Appendix~\ref{appendix:GADGET-preMerger-Starburst}. Since the sSFR of GADGET-3 and GADGET-4 is contaminated and thus cannot be representative of a galaxy's baseline sSFR, we excluded these two codes from our burst fraction calculation. We emphasise that the two codes still exhibit a starburst associated with the \textit{target} merger; however, we do not have a reliable means to quantify the strength of this burst. 

In the~100-Myr time window preceding the \textit{target} merger's infall, we chose the lowest sSFR as the baseline sSFR. This choice allows us to avoid small peaks of starburst that may be caused by very minor mergers ($\mu < 1:10$) that we cannot isolate. Fig.~\ref{fig:burst_fraction} demonstrates our method. In the top row, we show the sSFR history of the primary galaxy, excluding the contribution of stars that are at any time more bound to the secondary galaxy than the primary. The vertical gray band represents the~100-Myr window utilised to determine the baseline sSFR. The baseline sSFR is marked by the horizontal dashed coloured line and also written on the top left corner of the subplots in the top row. The baseline sSFR values indicate that all galaxies are star-forming (sSFR > $10^{-11}\,\text{yr}^{-1}$, a common threshold used in the literature, see \citealp{Katsianis+2021} and the references within). This baseline sSFR is used to calculate the baseline SFR and the baseline stellar mass, which are respectively shown as dashed coloured lines and dashed light gray lines in the bottom row of Fig.~\ref{fig:burst_fraction}. With this method, we can see that the predicted baseline stellar mass evolves at the same slope as right before the \textit{target} merger and has no sudden jumps, which matches our expectation of an isolated galaxy with no mergers. The formula for the burst fraction is,
\begin{equation}
    f_\text{sb} = \frac{M_{\bigstar,\text{actual}}  - M_{\bigstar,\text{baseline}}}{M_{\bigstar,\text{baseline}}}  \;\text{at}\;t=t_\text{sb},
    \label{eq:fsb}
\end{equation}
where $M_{\bigstar,\text{actual}}$ is the stellar mass of the main galaxy excluding any stars that are ever bound to the secondary galaxy, and $M_{\bigstar,\text{baseline}}$ is the predicted baseline stellar mass assuming a constant sSFR. The burst fraction is evaluated at $t_\text{sb}$, which is represented by the dotted line in the bottom row of Fig.~\ref{fig:burst_fraction}. For all codes but GEAR, $t_\text{sb}$ is chosen at the coalescence timestep. For GEAR, the galaxy slightly quenches following the \textit{target} merger, as evidenced by a gradual decline in the sSFR. Combined with the relatively long coalescence timescale, this decline in sSFR leads to the baseline stellar mass exceeding the actual stellar mass at about~1~Gyr after infall. Because GEAR's SFR does get enhanced by the merger, a negative burst fraction is not meaningful and uninformative. Therefore, instead of evaluating $f_\text{sb}$ at coalescence for GEAR, we evaluated it when the difference between the actual stellar mass and the baseline stellar mass is the largest. Physically, this corresponds to the moment when the GEAR galaxy's star formation is most enhanced (compared to a baseline SFR) before its sSFR diminishes. 

The $f_\text{sb}$ values for all \texttt{CosmoRun} codes but GADGET-3 and GADGET-4 are listed in the second row of Fig.~\ref{fig:burst_fraction}. Except for GEAR, all codes demonstrate a high burst fraction ($f_\text{sb} > 0.4$), as expected from a major merger \citep{Cox+2008}.
Among the considered codes, GIZMO and GEAR define the upper and lower extremes of $f_\text{sb}$. As GIZMO attains the highest SFR peak during the interaction, it correspondingly produces the largest $f_\text{sb}$, in which 99.5\% of the main galaxy's stellar mass is made up of merger-induced star formation. Indeed, this burst increases GIZMO's stellar mass by two orders of magnitude, going from $\approx 10^{8} \msun$ to $\approx 10^{10} \msun$ in about~500~Myr. On the contrary, the burst fraction of GEAR is only~0.263, considerably lower than that of other codes. This value is consistent with the fact that GEAR's SFR during the merger only increases slightly while fluctuating strongly due to the delayed cooling scheme. It is worth emphasizing that $f_\text{sb}$ quantifies relative star formation enhancement, not absolute enhancement. This relative definition means that even if the total stellar mass formed during the merger is similar between codes, $f_\text{sb}$ is larger for codes with lower pre-merger sSFR, and vice versa. 

After calculating the burst fractions, our next goal is to determine what may potentially explain their variation across the codes. Instead of comparing code to code, we seek to identify a common thread that determines the star formation enhancement strength regardless of the code being used. For example, \cite{Cox+2008} found that their burst efficiency becomes larger with increasing mass ratio and decreasing gas fraction of the primary galaxy; however, they only investigated two different feedback models (one with a stiff equation of state for star-forming gas and another that assumes isothermal star-forming gas). Fig.~\ref{fig:burst_fraction_correlation_with_PreMerger} displays the relationship between our calculated $f_\text{sb}$ and the system's gas fraction ($f_\text{gas}$) as well as the merger's total mass ratio ($\mu_\text{total}$, total mass is DM mass plus stellar mass plus total gas mass). We find a significant negative linear correlation ($R^{2} = 0.74$) between the burst fraction and the gas fraction of the merging system. This anticorrelation is indeed observed in multiple theoretical studies using binary galaxy merger simulations \citep[though their definitions of gas fraction are slightly different]{Cox+2008, Hopkins+2009, Scudder+2015, Fensch+2017a}, emphasizing its validity even in a cosmological simulation setting and over different code architectures and feedback models. A larger gas fraction means less stellar mass to generate interaction-driven gravitational torques to drive gas towards the galactic centre, leading to a smaller SFR enhancement \citep{Hopkins+2009}. \cite{Fensch+2017a} also reported that in galaxies with high gas fractions, gas inflows triggered by mergers are comparatively less important than pre-merger inflows, leading to a smaller relative SFR enhancement. We note that at sufficiently low gas fractions, the burst fraction may also be suppressed. \cite{Hopkins+2009} demonstrated that the threshold gas fraction below which the suppression occurs depends on the orbital parameter. Yet, a direct comparison with our results is not straightforward, because their analysis uses the cold disc gas fraction rather than the total gas fraction as in our analysis.  In Fig.~\ref{fig:burst_fraction_correlation_with_PreMerger}, we see a decreasing trend of $f_\text{sb}$ when $f_\text{gas}$ is smaller than~0.12. Nevertheless, given our small sample, it is difficult to conclude whether the trend is physical or a statistical fluctuation. 

We also find a very weak dependence of $f_\text{sb}$ on the merger's total mass ratio, where a higher total mass ratio leads to a stronger burst. This dependence is similarly observed in both previous observational \citep{Ellison+2008} and simulation works \citep{Cox+2008, Hopkins+2009, Hani+2020}. Besides $f_\text{gas}$ and $\mu_\text{total}$, we also checked other variables, such as the initial relative velocity magnitude, the angle between the galaxies' rotation and orbital plane (Table~\ref{tab:initial_state}), the periapsis and apoapsis distance, the merger's timescale, the DM mass ratio $\mu_\text{DM}$, the stellar mass ratio $\mu_\bigstar$, and the baryonic mass ratio $\mu_\text{baryon}$. None of those variables returned a significantly higher $R^{2}$. For conciseness, we do not include them in  Fig.~\ref{fig:burst_fraction_correlation_with_PreMerger}. It is important to note that since we compare different codes, the different feedback schemes may also play a role in regulating the strength of the burst and thus can undermine the physical relationship between the burst fraction and other variables.

\section{Discussion}
\label{sec:Discussion}

\subsection{Comparison with observations}
\label{sec:comparison_observations}

The changes in the SFR in our major merger at $z = 4.5$ across nine codes yield both similarities and differences when comparing with observational results. By using a sample of post-coalescence galaxies in the UNIONS survey identified by the MUMMI deep learning framework \citep{Ferreira+2024}, \cite{Ferreira+2025} calculated the SFR enhancements along the merger sequence and up to~1.5~Gyr after coalescence. They found that the SFR enhancement peaks within~500~Myr before or after coalescence, and quickly returns to the SFR of isolated galaxies at~$1\text{--}1.5$~Gyr after coalescence. In our simulation suite, ART-I, CHANGA, GADGET-3, GADGET-4, AREPO-T, and GIZMO (six out of nine codes) show a similar behaviour. Among these, CHANGA and AREPO-T have a much later SFR peak at about $0.5\text{--}0.7$ Gyr after coalescence, while others (codes with kinetic feedback) show an earlier peak slightly after the first periapsis. On the other hand, ENZO, RAMSES, and GEAR exhibit an overall increasing trend of SFR even~1.5~Gyr after coalescence with no clear decreasing trend. Nevertheless, \cite{Ferreira+2025} focused on local galaxies ($z < 0.3$), while our merger occurs at $z \approx 4.5$. At a higher redshift, the galaxies generally have a higher gas fraction and accretion rates, which can facilitate sustained star formation after the merging interaction. Indeed, even for codes with a pronounced merger-driven starburst such as GIZMO or GADGET-3, the galaxies are quickly rejuvenated with gas from accretion and maintain a high SFR instead of being quenched. 

It is important to note that the observational result is averaged from a large sample of galaxies, while this study only examines one merger in each of the nine simulations. Thus, the statistical averaging result may not well represent each individual galaxy merger, and one instance deviating from the mean does not necessarily imply that a simulation is incorrect. A bigger sample size of galaxy mergers in each simulation code is needed to make a more robust comparison between simulations and observations.

\subsection{Comparison with other theoretical works}

Seminal theoretical works by \cite{Barnes+1996} followed by \cite{Hopkins+2009} with major mergers of disc galaxies predicted that during the periapsides, the non-axisymmetric perturbation introduced by the secondary galaxy leads to the formation of a temporary stellar bar-like structure and a temporary gaseous bar-like structure. This response is non-axisymmetric and is not similar to a morphological bar in barred spiral galaxies. Due to the collisional nature of the gas and the collisionless dynamics of the stellar component, the stellar bar-like structure typically trails the gas bar-like structure by a small angular offset. As a result, the internal stellar structure exerts a gravitational torque on the gas structure, removing its angular momentum and driving gas inflow toward the central regions, causing a starburst. In their work, the SFR peaks during periapsides and quickly goes down. \cite{Hopkins+2009} also claimed that the simulation feedback prescriptions should only change the intensity of the starburst and not the general dynamical processes. Nonetheless, they only investigated simulations that employ kinetic feedback \citep[from][]{Mihos+1994} or galactic wind \citep[from][]{Springel+2003a}, which is a form of propagating kinetic energy out during supernovae. Therefore, the findings of \cite{Hopkins+2009} are not representative of mergers in all \texttt{CosmoRun} codes, but rather only those with kinetic feedback. Indeed, in our case, the starburst during periapsis is only prominent in ART-I, GADGET-3, GADGET-4, and GIZMO, which are codes implemented with kinetic feedback. Some codes using thermal feedback without kinetic feedback (ENZO, AREPO-T, CHANGA) do also exhibit a modest "burst" in SFR during the first periapsis (around the middle of the first passage stage, yellow column). Yet, this increase is only a $2\text{--}3$ $\msun/\text{yr}$ increment, which is smaller than the global increase during the merging interaction (around $10$ $\msun/\text{yr}$ in ENZO and CHANGA and around $25$ $\msun/\text{yr}$ in AREPO-T). This small SFR increase shows that the dynamical process of inducing a starburst from \cite{Hopkins+2009} is still relevant, yet the absence of kinetic feedback suppresses its effect. 

While our study examines the effects of supernova feedback by comparing codes, individual code groups have addressed the same question by employing different feedback strategies within their own code. Several of those studies yield consistency with our results. \cite{Chaikin+2023} tested the relative importance of kinetic feedback and thermal feedback using the particle-based code SWIFT and found that a more kinetic-dominated model leads to a weaker galactic wind (lower wind mass loading factor) and a greater central gas surface density. In their isolated disc galaxy simulations using RAMSES, \cite{Rosdahl+2017} showed that a delayed cooling prescription is the most efficient scheme at smoothing out the gas distribution and thickening the disc, which is similar to what we observe in Fig.~\ref{fig:gas_density_distance_plot}. They also demonstrated that when using only kinetic feedback, a very thin disc plane and well-defined, relatively dense spiral filaments appear. Furthermore, \cite{Nunez-Castineyra+2020} found in RAMSES that, compared to mechanical feedback, a delayed cooling feedback scheme combined with the Kennicutt-Schmidt law star formation scheme (with constant star formation efficiency) results in a well-extended and diffused gas disc and a less concentrated bulge. We further emphasise that some studies caution that the magnitude of these discrepancies is resolution-dependent, and different feedback schemes may converge when the simulation has a sufficient mass resolution \citep{Hopkins+2018, Smith+2018}. It is important to note that these controlled comparisons were carried out for isolated discs. For galaxy mergers, their complex dynamical processes can alter these feedback trends.

\subsection{Caveats}
\label{sec:caveat}

We identify three caveats of this work. Firstly, we assume that the very minor mergers ($1:100 <\mu < 1:10$, anything less than~1:100 can be considered part of accretion from filament) around the \textit{target} merger (as shown in Fig.~\ref{fig:merger_tree}) contribute negligibly to changes in SFR compared to the \textit{target} merger. This assumption is motivated by previous studies that showed the small importance of these mergers in SFR enhancement \citep{Cox+2008, Lotz+2010}. Nonetheless, we see an example where the assumption is not entirely valid. In Appendix~\ref{appendix:GADGET-preMerger-Starburst}, we discuss a hypothesis that a merger with a DM mass ratio of~3:100 may be responsible for triggering a strong starburst in GADGET-3 and GADGET-4. We find that, among all \texttt{CosmoRun} codes, GADGET-3 and GADGET-4 have very low average gas collapse times before this very minor merger, and the merger may perturb the gas to further collapse and cause this starburst. Disentangling the effects of very minor mergers from a simultaneous major merger is inherently challenging. Nevertheless, there are still several considerations that support our assumption. The strong star formation response from very minor mergers does not happen in other codes besides GADGET-3 and GADGET-4. In addition, there are fewer than~12 very minor mergers (most of which have a mass ratio of only~1:100 to~3:100) entering the main galaxy's boundary during the \textit{target} merger's timescale, while the \textit{target} merger has a very high mass ratio~(1:2 to 1:1). Therefore, we maintain our assumption that very minor mergers play a negligible role in driving the observed global trends compared to the \textit{target} merger. 

Secondly, we neglected the stochastic nature of galaxy simulations in our work. Using simulations of isolated dwarf galaxies, \cite{Keller+2019} demonstrated that tiny numerical perturbations introduced by floating-point round-off, random number generators, and seemingly trivial differences in algorithmic behaviour can lead to non-trivial differences in simulated galaxy properties. Such differences grow substantially during major mergers, for which the stellar mass of the merger remnant varies by a factor of two between otherwise identical runs. This run-to-run chaotic behaviour can further complicate our comparison, as any detected divergence between the merger remnants may partly reflect the numerical stochasticity rather than the feedback prescription choices.

The third caveat is the study's small sample size. The outcome of galaxy mergers can differ widely depending on the merger's physical parameters, such as its orbital configuration, mass ratio, progenitor galaxies' stellar mass, etc. Thus, even though our finding indicates that each simulation's stellar feedback scheme influences the merger's response in a specific way, this finding may not necessarily be universal for all mergers. A bigger sample of galaxy mergers with varying sets of physical parameters and high convergence on the merger's parameters between the codes (like with the \textit{target} merger) is therefore needed to holistically investigate the extent of our conclusions. A bigger sample can also help test whether the simulation's stellar feedback scheme can bias our prediction of galaxy assembly when applied to a larger box size. In addition to the merger's physical parameters, changing the total feedback energy injected or the relative contributions of the individual feedback channels (e.g., between thermal and kinetic feedback) can also influence the results. Nevertheless, even with the sample size of one, our study still highlights the dependence of the outcome of galaxy mergers on the stellar feedback subgrid models.

\section{Conclusion}
\label{sec:conclusion}

In this paper, we use the AGORA \texttt{CosmoRun} suite of nine cosmological zoom-in hydrodynamic simulations of a Milky Way-mass halo to examine the effect of a major galaxy merger at $z \approx 4.5$ on the star formation activity of the main galaxy. The simulations in the suite are carefully calibrated to each other so that the discrepancies in the merger remnant's properties are mostly attributed to the stellar feedback schemes and the code architectures. The halo tree is built using \textsc{HASKAP PIE}, a state-of-the-art halo finder that defines halos by only bound particles and in a non-spherical fashion. Our key findings from the cross-code comparison of the major merger are as follows:

\begin{itemize}
    \item The SFR evolution during the merging interaction is governed by the type of stellar feedback used in each code group. Codes that employ kinetic feedback in their model experience a starburst starting from the first periapsis, followed by a decline in SFR before the two galaxies coalesce. Codes using thermal feedback without kinetic feedback show an increasing trend of SFR that continues well into the post-coalescence stage. Lastly, the use of the delayed cooling scheme or the radiation pressure scheme creates a short-timescale, bursty SFR with small amplitudes throughout the interaction (Figs.~\ref{fig:mass_sfr}). We do not find any systematic dependence of the SFR patterns on the choice of hydrodynamics solver (grid-based, particle-based, hybrid) or on the merger's physical parameters.

    \item Examining the gas properties reveals that kinetic feedback helps drive gas to high density, high concentration, and low temperature at the galactic centre earlier in the merging interaction. When using only thermal feedback, gas cools and reaches high central density only after coalescence. When adding a delayed cooling or superbubble scheme to thermal feedback (without kinetic feedback), a dense central gas concentration is suppressed, resulting in a more extended gas distribution. We also find that the total gas reservoir shows no clear correlation with the SFR patterns (Figs.~\ref{fig:mass_sfr}, \ref{fig:gas_phase_plot}, and  \ref{fig:gas_density_distance_plot}). 

    \item By tracing the trajectories and thermal histories of gas particles that form stars during the merger in particle-based codes, we demonstrate that kinetic feedback (GADGET-3, GADGET-4, GIZMO) facilitates a more rapid accretion of gas particles from the secondary galaxy onto the primary galaxy during the first periapsis, consequently driving an early starburst. Thermal feedback, combined with superbubble (CHANGA) and delayed-cooling (GEAR) schemes, inhibits efficient gas cooling and gas accretion towards the primary galaxy's centre, thereby producing a more subdued, extended star formation response. (Figs.~\ref{fig:track_gas_particles_turning_into_stars_fromsecondaryHalo} and \ref{fig:infalling_momentum})

    \item We introduce a novel way to calculate the burst fraction (the degree of star formation enhancement) of a galaxy merger in cosmological simulations. This method does not require a controlled group of isolated galaxies, yet it needs a quiet SFR history right before the merger event. We find a negative correlation between our calculated burst fraction and the pre-merger gas fraction of the progenitor galaxies, which agrees with previous studies indicating that more gas-rich progenitors suppress merger-induced starbursts. A relatively weak correlation between the burst fraction and the merger's total mass ratio is also observed. These correlations are independent of the employed stellar feedback model. (Figs.~\ref{fig:burst_fraction} and ~\ref{fig:burst_fraction_correlation_with_PreMerger})

    \item Most of our simulated mergers are relatively consistent with observational results regarding the timing of the merger-driven star formation. (Section~\ref{sec:comparison_observations})
    
\end{itemize}

This study emphasises the need for caution when interpreting the properties of simulated galaxy mergers, as the simulation's stellar feedback scheme may prejudice the mergers' outcomes to a specific response in star formation. A suite of calibrated cosmological simulations with a larger volume is therefore needed to further test whether the choice of stellar feedback scheme can systematically bias the predicted star formation response for other merger configurations, or more broadly, the predicted contribution of mergers to cosmic star formation.

\section*{ACKNOWLEDGEMENTS}

We thank all of our colleagues in the AGORA Collaboration for their collaborative and supportive spirit, which has allowed the collaboration to remain strong as a platform to foster and launch multiple science-oriented comparison efforts. THN acknowledges support from the National Center for Supercomputing Applications Center and its Center for Astrophysical Surveys. THN and KSSB acknowledge the University of Illinois at Urbana-Champaign for their continued support, and the support of the Delta Supercomputer as well as the ACCESS program for computing grants PHYS240175 and PHY250173.
R.R.C.acknowledges financial support from the Spanish Ministry of Science and Innovation through the research grants: PID2021-123417OBI00, funded by MCIN/AEI/10.13039/501100011033/FEDER, EU; PCI2022-135023-2, funded by MCIN/AEI/10.13039/ 501100011033; the EU “NextGenerationEU” / PRTR; and PID2024-157374OBI00, funded by MICIU/AEI/10.13039/501100011033/FEDER, EU; and the IND2022/TIC-23643 project funded by Comunidad de Madrid. 
J.-H.K.’s work was supported by the National Research Foundation of Korea (NRF) grant funded by the Korea government (MSIT; No. 2022M3K3A1093827 and No. 2023R1A2C1003244). His work was also supported by the National Institute of Supercomputing and Network/ Korea Institute of Science and Technology Information with supercomputing resources including technical support, grants KSC-2022-CRE-0355 and KSC-2024-CRE-0232. His work was also supported by the GlobalLAMP Program of the NRF grant funded by the Ministry of Education (No. RS-2023-00301976). 
KN acknowledges support from JSPS KAKENHI grant 20H00180, 24H00002, 24H00241, JP25K01032, and the JSPS International Leading Research (ILR) project, JP22K21349.
KN also acknowledges support from the Kavli IPMU, the World Premier Research Center Initiative (WPI), UTIAS, the University of Tokyo.  
DC is supported by research grant  PID2024-156100NB-C21 financed by MICIU/AEI /10.13039/501100011033 / FEDER, EU., and the research grant
CNS2024-154550 funded by MI-CIU/AEI/10.13039/501100011033.  
HV was supported by a grant UNAM-PAPIIT-IN111425.

\section*{DATA AVAILABILITY}
The AGORA \texttt{CosmoRun} raw simulation snapshots from Papers~\citetalias{Roca-Fabrega+2021} and \citetalias{Roca-Fabrega+2024} are publicly available at \url{https://flathub.flatironinstitute.org/agora}. The analysis codes and the simulation metadata underlying this project can be shared on reasonable request to the corresponding authors and the AGORA collaboration.

%

\vspace{5mm}



\appendix

\section{Evaluating the coalescence criteria}
\label{appendix:evaluating_massratio_coalescence}

\begin{figure*}[tbh]
    \centering
    \includegraphics[width=0.9\linewidth]{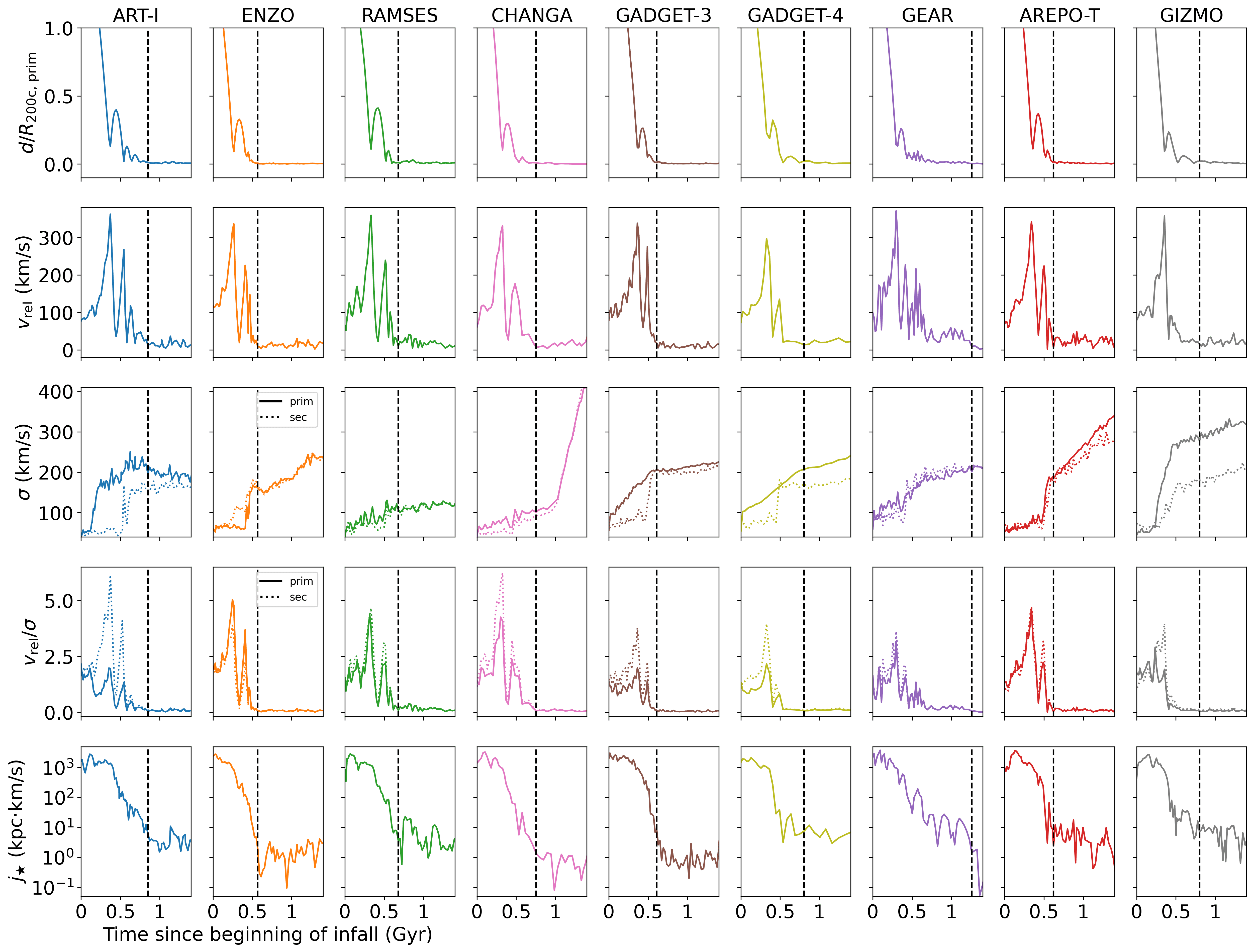}
    \caption{The time evolution of the variables used to define coalescence in Equation~\ref{eq:coalescence_condition}. From top to bottom, the rows show the ratio between the distance between the two galaxies and the primary galaxy's $R_{200c}$ radius, the relative velocity between the two galaxies, the velocity dispersion of the primary (solid lines) and secondary galaxy (dotted lines), the ratio between the relative velocity magnitude and the velocity dispersion of each progenitor, and the orbital angular momentum of the secondary galaxy with respect to the primary galaxy. All variables are computed using the galaxies' stellar cores. At our determined coalescence timestep (vertical black dashed lines), the relative specific angular momentum $j_\star$ has diminished to less than~1\% of its initial value, which is consistent with the commonly used definition of the end of a merger in the literature.}
    \label{fig:d_v_j}
\end{figure*}

As we employed a new approach to define coalescence, in this appendix, we discuss the details of the variables used in our definition and assess the consistency of our criteria with those used in previous work. Our three criteria to determine the coalescence timestep (Equation~\ref{eq:coalescence_condition}) depend on the distance between the two galaxies, the primary halo's radius, the relative velocity between the two galaxies, and the velocity dispersion of the two galaxies' stellar cores. Fig.~\ref{fig:d_v_j} shows the time evolution of each of those variables during the \textit{target} merger. $R_{200c}$ is chosen to represent the halo radius, where the overdensity of 200 carries no special physical significance with respect to the coalescence criterion. Alternative overdensity radii may be adopted, provided the corresponding normalisations are adjusted accordingly. Because we normalised velocity and distance in the formulas, these coalescence criteria can be applied for mergers at other mass scales and mass ratios. Also, the normalisation by the velocity dispersion helps take into account the situation where the stellar cores are potentially tidally disrupted.

Another common way to define the end of a galaxy merger is when the satellite galaxy loses all or a large fraction of its specific angular momentum relative to the host and/or its bound particles \citep[e.g.]{Boylan-Kolchin+2008, McCavana+2012, Villalobos+2013}. Nevertheless, the angular-momentum condition does not account for a situation in which the secondary galaxy comes in a head-on collision with the primary galaxy, where its relative specific angular momentum is zero during the whole interaction because the velocity and positional vectors are in the same direction. To avoid this edge case, we instead put constraints on the distance and the relative centre-of-mass velocity. When both the distance and the relative speed are zero, the angular momentum will also be zero, allowing us to skip the constraint on the angular momentum. 

To evaluate our coalescence definition with the relative angular momentum, we show in the last row of Fig.~\ref{fig:d_v_j} the evolution of the specific angular momentum of the secondary galaxy relative to the primary galaxy ($j_\star$). In all codes, $j_\star$ decreases continuously with very small fluctuations throughout the interaction, which is expected as the secondary galaxy loses angular momentum through dynamical friction during its infall. Even though angular momentum is not explicitly incorporated in our coalescence criteria, at our determined coalescence timestep, $j_\star$ is still less than~1\% of its value at the timestep when the secondary galaxy crosses the primary galaxy's virial radius. We also checked the number of bound star particles of the secondary galaxy, and we confirm that at our chosen coalescence timestep, the secondary galaxy has already lost all of its bound particles or lost them to the primary galaxy (the stars become more bound to the primary galaxy than to the secondary galaxy). To sum up, our coalescence criteria are consistent with the angular momentum loss and bound particle loss definitions of coalescence in the literature. 

Fig.~\ref{fig:d_v_j} also shows why the \textit{target} merger in GEAR takes significantly more time to coalesce compared to other codes (right subplot of Fig.~\ref{fig:merger_timing_and_stages}). While other codes only show two to three periapsis passages before the system relaxes, the two stellar cores in GEAR undergo at least six resolved passages. Moreover, the apoapses between the second and sixth passages are relatively similar to each other, demonstrating that the secondary galaxy's stellar core does not effectively lose angular momentum and maintains a relatively stable two-body orbit with the primary galaxy's stellar core. This relatively stable orbit is also reflected in the high relative velocity (second-row plot) and the high relative angular momentum (fifth-row plot) during these periapsis passages. This prolonged orbit is likely caused by GEAR's delayed cooling scheme in combination with its high gas fraction, which creates a high-pressure-supported system and consequently suppresses coalescence.

\section{GADGET-3 and GADGET-4's starburst before the \textit{target} merger}
\label{appendix:GADGET-preMerger-Starburst}

\begin{figure}[tbh]
    \centering
    \includegraphics[width=0.9\columnwidth]{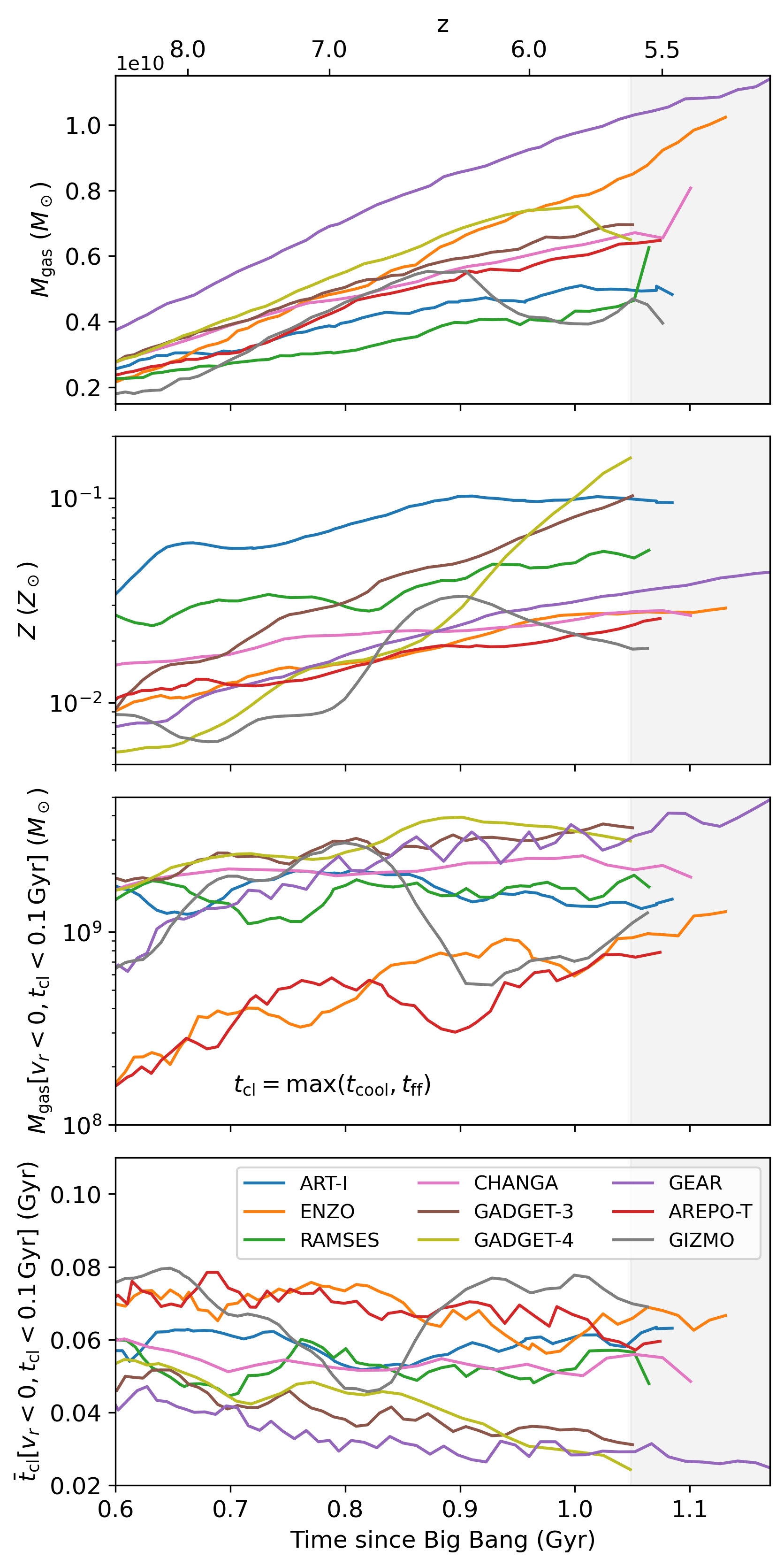}
    \caption{The gas mass (first row), gas metallicity (second row), fast-collapsing gas mass (third row), and average collapse time of the fast-collapsing gas (fourth row) within the main halo's convex hull before the \textit{target} merger. The vertical grey band represents the range of the \textit{target} merger's $t_\text{start}$ across the nine codes. A fast-collapsing gas element is defined as one with a negative radial velocity and a collapse time (taken as the greater of the free-fall time and the cooling time) smaller than~0.1 Gyr. Apart from GEAR, GADGET-3 and GADGET-4 have the largest amount of fast-collapsing gas and the shortest average collapsing time.}   
    \label{fig:GasMetallicity_and_CoolingTime_PreMerger}
\end{figure}

\begin{figure*}[tbh]
    \centering
    \includegraphics[width=\linewidth]{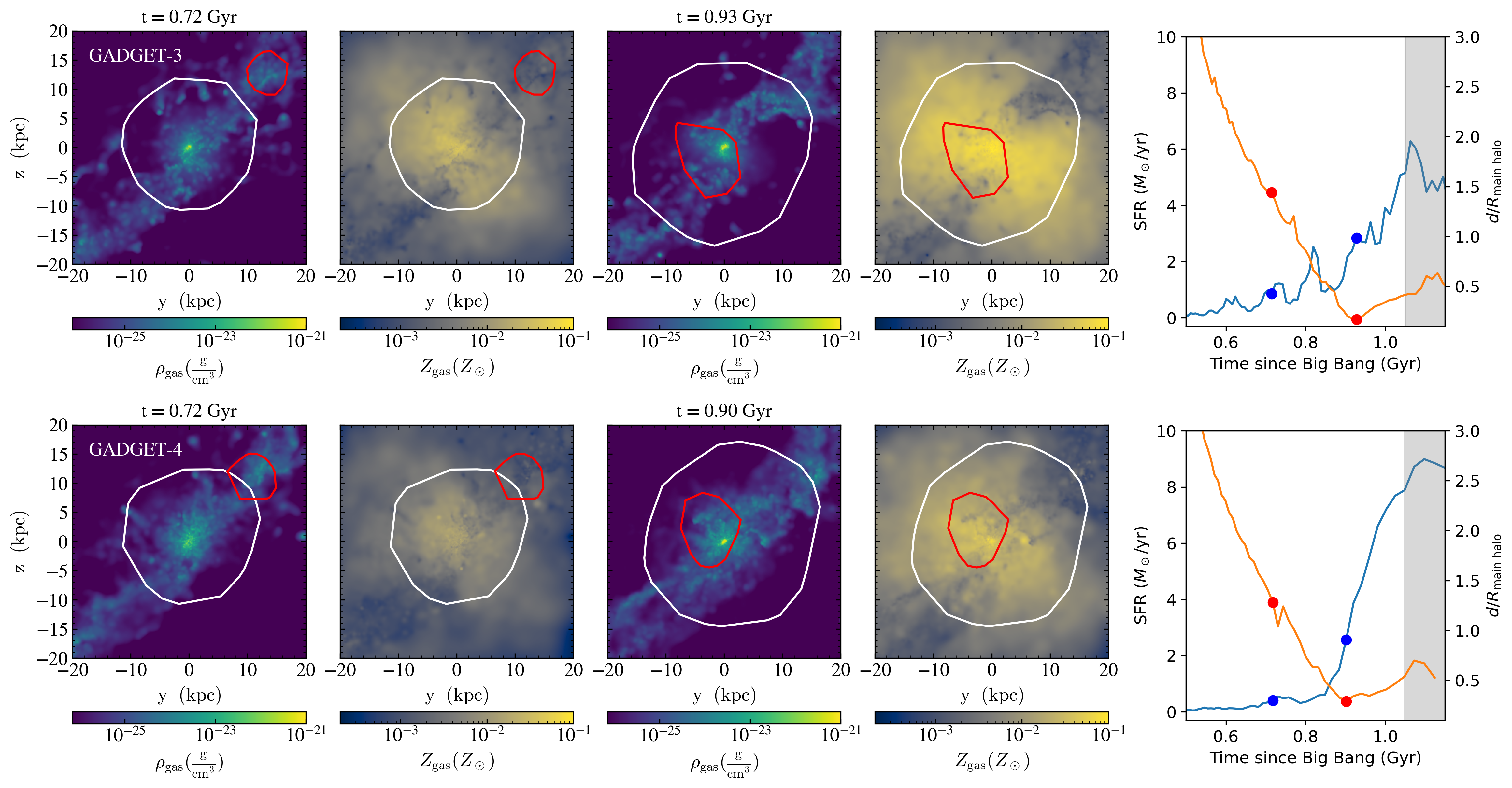}
    \caption{The gas density and gas metallicity projection plots, overlaid with \textsc{HASKAP PIE}'s non-spherical halos, showing the very minor merger (mass ratio $\approx$ 3:100, indicated by the red boundary) that may cause the $z \approx 6$ starburst of GADGET-3 (top row) and GADGET-4 (bottom row) before the \textit{target} merger. The white boundary marks the main halo. Two timesteps are shown for each code: one approximately at the infall and one at the first periapsis. The rightmost subplots show the merger's trajectory (orange) and the SFR evolution of the main halo (blue), and the scatter points locate the timesteps of the projection plots. The vertical grey band represents the beginning of the \textit{target} merger. The timing of this minor merger coincides with the collapse of central gas and the strong starburst at $z \approx 6$ in the two GADGET codes.}
    \label{fig:GADGET-preMergerStarburst}
\end{figure*}

In Fig.~\ref{fig:mass_sfr_evolution_premerger}, we identify a strong starburst occurring in GADGET-3 and GADGET-4 at $z \approx 6$, which is before the major merger at $z \approx 4.5$. This burst substantially augments the stellar mass of the main galaxy, allowing GADGET-3 and GADGET-4 to have the most massive galaxies among all codes at $z \approx 6$, despite GADGET-4 being one of the least massive systems prior to the burst. Not only does it affect the stellar mass, but this burst at $z \approx 6$ also makes the main galaxy in GADGET-3 and GADGET-4 become more compact and form a stellar disc (see fig.4 of Paper~\citetalias{Jung+2025a}). We identify two mechanisms to explain this burst. 

First, gas in the main halo of GADGET-3 and GADGET-4 has a lower collapse time than in other codes. The collapse time ($t_\text{cl}$) is computed as the larger value between the cooling time and the free-fall time of each gas element. To estimate the amount of gas collapsing to the galactic centre, we imposed two additional requirements: a gas element needs to have an infalling motion to the galactic centre (radial velocity $v_{r} < 0$), and the collapse time needs to be smaller than~0.1 Gyr. We refer to the gas elements satisfying these two conditions as fast-collapsing gas. The third row of Fig.~\ref{fig:GasMetallicity_and_CoolingTime_PreMerger} shows the time evolution of the main halo's fast-collapsing gas mass, and the fourth row shows the mass-weighted average collapse time of all the fast-collapsing gas elements. Among all nine codes but GEAR, even several hundred Myr before the $z \approx 6$ burst, GADGET-3 and GADGET-4 have the largest amount of fast-collapsing gas and the lowest average gas collapse time. The significantly lower $t_\text{cl}$ can explain why the starburst occurs in GADGET-3 and GADGET-4. We can also observe that after the burst starts, the ISM gets enriched with metals from supernovae (second row of Fig.~\ref{fig:GasMetallicity_and_CoolingTime_PreMerger}), leading to an even lower cooling time and lower collapse time. A low $t_\text{cl}$ requires high gas density, low gas temperature, and high gas metallicity. According to the second row of Fig.~\ref{fig:GasMetallicity_and_CoolingTime_PreMerger}, the main halo in GADGET-3 and GADGET-4 exhibits a rapid metal enrichment and becomes one of the most metal-rich halos among all nine codes, especially during and after the $z \approx 6$ starburst. Nonetheless, gas metallicity alone cannot explain the low cooling time, as ART-I and RAMSES also show high gas metallicity. It is an interplay of gas metallicity, density, and temperature regulated by GADGET-3 and GADGET-4's stellar feedback schemes that causes $t_\text{cl}$ to be low, leading to a more effective collapse and subsequently a strong starburst. Lastly, although GEAR similarly produces a substantial amount of fast-collapsing gas and comparably short average $t_\text{cl}$, its delayed-cooling scheme numerically suppresses radiative cooling, making our estimated $t_\text{cl}$ unrepresentative of the gas behaviour. As seen in Fig.~\ref{fig:gas_phase_plot}, the strong delayed cooling scheme of GEAR leads to a massive accumulation of very hot, dense gas. As the cooling time is scaled with $\rho^{-2}$ and only with $T$, a sufficiently high gas density can still yield a short computed cooling time even when the gas temperature is high. Thus, the computed collapse time is also short, as illustrated in the fourth row of Fig.~\ref{fig:GasMetallicity_and_CoolingTime_PreMerger}. Despite that, the effective cooling time (and hence the effective collapse time) is still infinite because of the numerical delay, which consequently prevents gas from collapsing to fuel a strong starburst. Though GADGET-3 is also implemented with a delayed cooling prescription, as discussed in Section~\ref{subsect:delayedcooling_gasproperties}, its specific implementation allows warm-hot dense gas to have a much shorter delayed cooling time, which facilitates quicker cooling and has less effect on $t_\text{cl}$ compared to GEAR. 

The second potential mechanism to explain this starburst of the two GADGET codes is that it is triggered by a very minor merger occurring at $z \approx 6\text{--}7$, which is approximately the same time the burst happens. In both of the codes, this minor merger has a DM mass ratio of 0.032 and begins at $z \approx 7.1$ ($t \approx 770$ Myr after the Big Bang). Fig.~\ref{fig:GADGET-preMergerStarburst} shows the location of the main halo (white border) and the secondary halo (red border) on top of the gas projection plot at the beginning of the infall and at the first periapsis of this very minor merger. The rightmost column shows the SFR time evolution and the normalised distance between the main galaxy and this merger's secondary galaxy. We can observe that the moment the SFR gets considerably enhanced coincides with the very minor merger's first periapsis. The gas projection also shows that after the first periapsis, gas collapses into the galactic centre, creating a much denser and more compact central region and giving rise to the starburst. Because the merger's mass ratio is only 0.03, the total gas mass barely increases, as seen in the first row of Fig.~\ref{fig:GasMetallicity_and_CoolingTime_PreMerger}. Therefore, the $z = 6$ starbursts in GADGET-3 and GADGET-4 are likely caused by the dynamics of the very minor merger that stimulates the collapse of gas (for example, by decreasing the angular momentum of the central gas) rather than a mere increment of gas inflow. We note that this very minor merger is also present in the other \texttt{CosmoRun} codes; however, further analysis is required to explain why the merger does not trigger a burst in them.

This example of GADGET-3 and GADGET-4 shows that a strong star formation episode and a galaxy compaction event can be triggered by a sufficiently low cooling time and/or by a very minor merger, instead of a merger with a large mass ratio $\geq 1:10$. This finding highlights the complexity of starburst pathways and provides a nuanced perspective on the conventional view that strong starbursts are primarily associated with major mergers, emphasizing the importance of thermodynamic conditions and smaller mergers \citep{Mihos+1994a}. Further work is needed to investigate and disentangle which one of our hypotheses is the dominant channel leading to the starburst, and whether this type of starburst happens in other galaxies in \texttt{CosmoRun}. 

\section{CHANGA delayed starburst}
\label{subsect:CHANGA_delayed_starburst}

\begin{figure}[tbh]
    \centering
    \includegraphics[width=0.9\columnwidth]{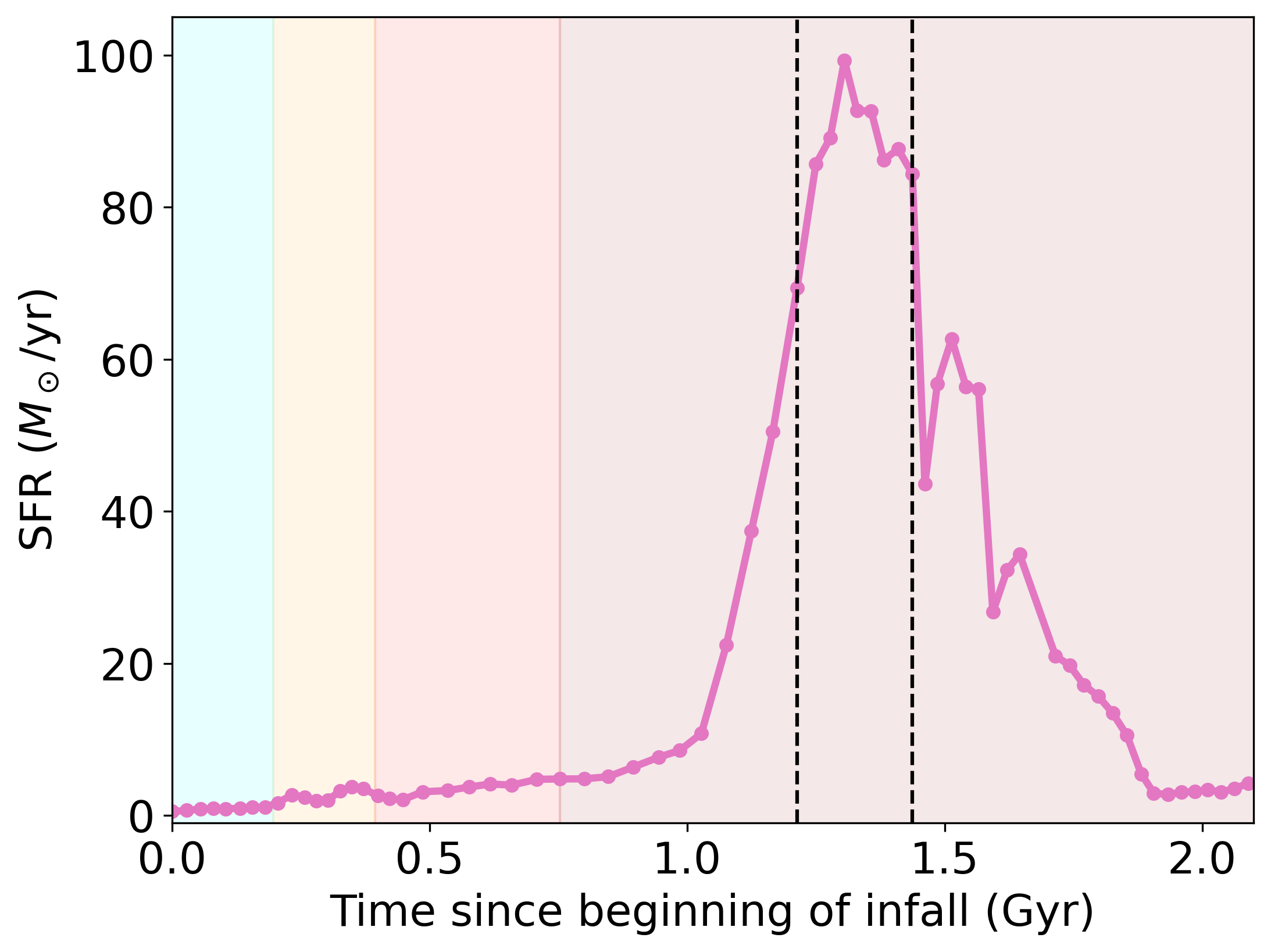}
    \caption{The CHANGA's delayed starburst that happens more than~500 Myr after coalescence. Even though this burst is delayed, it is the largest burst due to the \textit{target} merger across all simulations. The two dashed black lines mark the burst period that will be used to plot Fig.~\ref{fig:Trace_Gas_CHANGA_delayedStarburst}.}
    \label{fig:CHANGA_SFR_delayed_starburst}
\end{figure}

\begin{figure*}[tbh]
    \centering
    \includegraphics[width=0.95\linewidth]{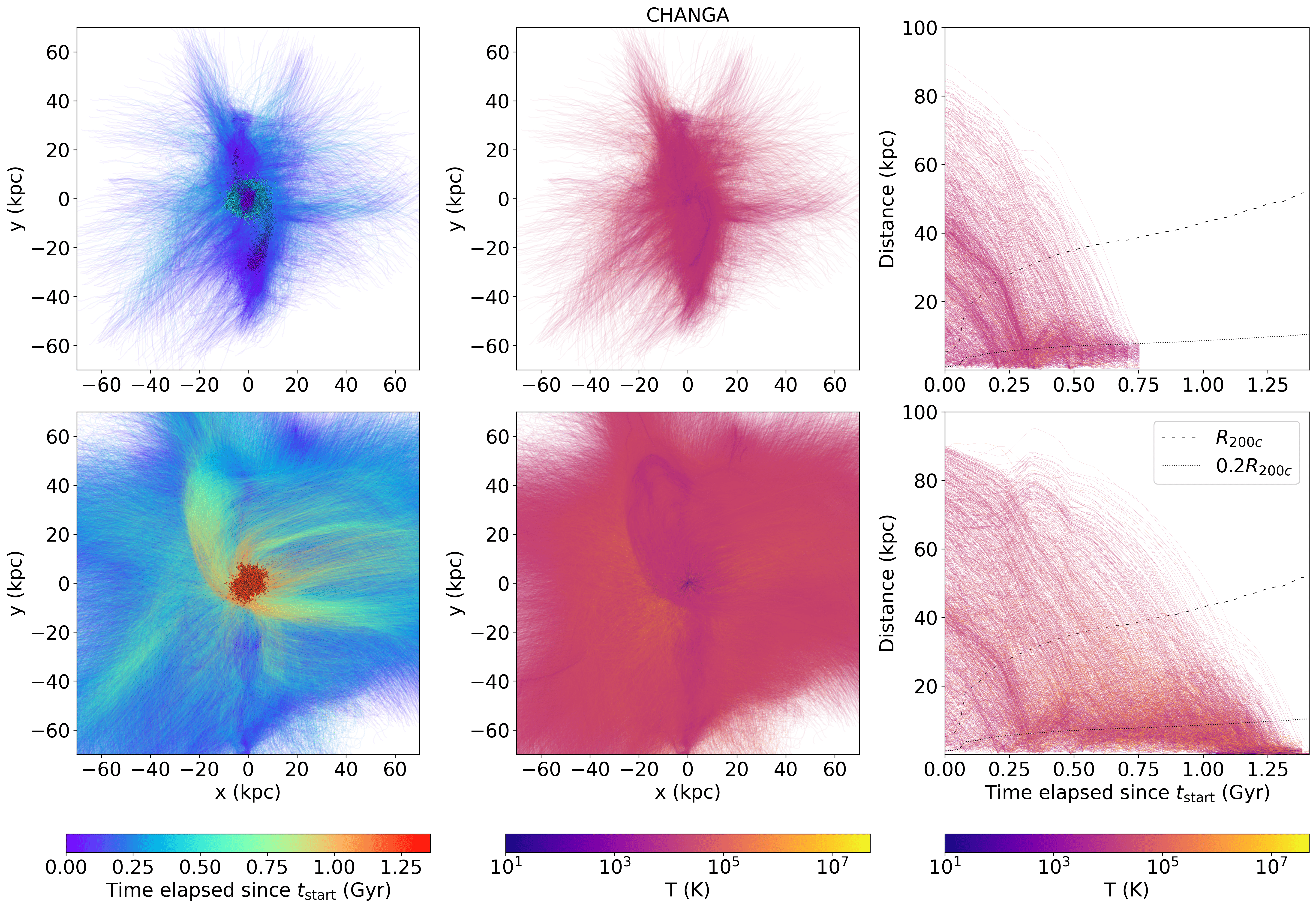}
    \caption{The trajectories and thermal evolution of gas particles that form stars during the merger timescale (top row) and during the delayed starburst (bottom row) in CHANGA. The left and right columns show the same type of plots as the top and middle rows of Fig.~\ref{fig:track_gas_particles_turning_into_stars_fromsecondaryHalo}. We plot the trajectories of all gas particles that are converted into stars, including gas originating from the secondary galaxy, minor halo mergers, and accretion. In the right column, we randomly selected~2000 gas particles to better display the time evolution of the gas particles' distance to the galactic centre. The dashed lines and the dotted lines represent the main halo's $R_{200c}$ and $0.2R_{200c}$, respectively. Gas that forms stars prior to coalescence accretes onto the main galaxy with minimal heating, whereas gas contributing to the delayed starburst is first heated by feedback during the merger and expelled away from the galaxy. This outflow then leads to fountain accretion and fuels the delayed starburst in CHANGA.}
    \label{fig:Trace_Gas_CHANGA_delayedStarburst}
\end{figure*}

As illustrated in Fig.~\ref{fig:mass_sfr} and shown in previous works \citep{Cox+2008, Lotz+2008, HungHongZhaoLing+2016}, the peak of a merger-induced starburst typically occurs within the merger timescale or around coalescence. Nevertheless, in \texttt{CosmoRun}, we find an exception with the \textit{target} merger of CHANGA. Even though the SFR gradually rises throughout the merging interaction, CHANGA's most intense starburst does not take place within the \textit{target} merger's timescale, but approximately~$550\,\text{Myr}$ after coalescence. Fig.~\ref{fig:CHANGA_SFR_delayed_starburst} demonstrates that this pronounced SFR burst emerges well into the post-coalescence stage, reaching a peak value roughly~40 times higher than the time-average SFR during the \textit{target} merger timescale. In fact, this SFR burst of CHANGA surpasses all other \textit{target} mergers' bursts in \texttt{CosmoRun} and attains a peak value approximately three times greater than the burst observed in GIZMO (the largest burst within the \textit{target} merger's timescale, see Fig~\ref{fig:mass_sfr}).

While studying morphological compaction, Paper~\citetalias{Jung+2025a} (see their fig. 12) also mentioned this behaviour and explained that the delayed starburst was caused by gas re-accreting onto the main galaxy after being expelled due to stellar feedback during the \textit{target} merger. We further examined this behaviour by looking at the trajectories of the gas particles that turn into stars during the \textit{target} merger's timescale and during the delayed starburst. We first identified the gas particles that disappear during the examined period and are located within the region covering both galaxies or within~0.2$R_\text{200c}$ after the two galaxies coalesce. Next, we tracked those gas particles back in time. It is important to note that, unlike Fig.~\ref{fig:track_gas_particles_turning_into_stars_fromsecondaryHalo}, which isolates only the gas particles from the secondary galaxy, we included in Fig.~\ref{fig:Trace_Gas_CHANGA_delayedStarburst} all the gas that contributes to star formation during the interaction, including gas from the secondary galaxy, from smaller halo mergers, and from accretion from the cosmic filament. The top row of Fig.~\ref{fig:Trace_Gas_CHANGA_delayedStarburst} shows the stars that formed between $t_\text{start}$ and $t_\text{cls}$, while the bottom row shows the stars that formed during the delayed starburst period (marked by the two vertical dashed lines in Fig.~\ref{fig:CHANGA_SFR_delayed_starburst}). The left column shows the gas trajectories coloured by time and the location of the stars when they first form, and the middle column shows the gas trajectories coloured by their temperature. In the right column, for a random subset of 2000 gas particles, we plot their distance to the primary galaxy's centre as a function of time, with each line coloured by its temperature evolution. When randomly selecting gas particles that form stars during the delayed starburst (bottom row), we excluded those that are initially too distant ($> 90$~kpc) from the main galaxy, allowing a clearer comparison with the trajectories of gas particles that form stars during the merger timescale (top row).

We see significant differences in the trajectory and thermal evolution of gas particles that form stars prior to coalescence and during the delayed starburst. During the \textit{target} merger, cold gas inflow (from the secondary galaxy and accretion) falls onto the central galaxy and then forms stars with little heating. On the other hand, the gas that contributes to the delayed starburst has a much more extended trajectory. These gas particles also flow in from the secondary galaxy and accretion; however, instead of collapsing to form stars, they receive thermal feedback from stars that form during the merger and get ejected into the CGM. Comparing the bottom-right and top-right subplots of Fig.~\ref{fig:Trace_Gas_CHANGA_delayedStarburst}, we find that gas particles forming stars during the delayed burst are pushed further out during their infall toward the galactic centre, producing various noticeable "loops" that extend beyond the ISM ($\approx 0.2R_{200c}$) into the CGM ($\approx 0.2R_{200c} < \text{distance} < R_{200c}$). Moreover, the gas temperature between radii of~5 and~30~kpc is higher in the bottom-right subplot, reflecting the heating feedback that drives the outflow. CHANGA employs the superbubble feedback, which effectively combines feedback energy from multiple star particles into a single hot "bubble" governed by thermal conduction. The prescription supports multiphase gas elements (hot and cold phases in pressure equilibrium, with a threshold between two phases being at $10^{5} \text{K}$) during the superbubble phase, helping avoid overcooling in unresolved gas elements. Thus, the feedback scheme allows resolved hot gas to continue to gain mass from nearby cold gas, leading to a higher amount of hot gas than only depositing thermal feedback in surrounding gas cells \citep{Keller+2014}. As a result, the superbubble feedback drives much more outflow than using traditional blastwave feedback.  

The supernova-driven, more metal-rich outflow gas then mixes with the hot coronal CGM gas, leading to a reduction of the cooling time of the hot gas and facilitating subsequent accretion, a mechanism known as fountain accretion \citep{Fraternali+2017}. This fountain accretion starts to happen around~1.1 Gyr after the infall, causing the delayed starburst in CHANGA. This process is also demonstrated by the thermal evolution shown in the bottom middle and bottom right subplots of Fig.~\ref{fig:Trace_Gas_CHANGA_delayedStarburst}, where we observe that the gas is hot ($> 10^6 K$) after being expelled out of the galaxy, yet it cools down again after reaching the CGM and then accretes back onto the galaxy. Furthermore, the large amount of gas from this "delayed accretion" forms a prominent stellar disc for the main galaxy (see fig.~4 of Paper~\citetalias{Jung+2025a}). This is in contrast to a more spherical distribution of stars formed during the merger timescale (details in Paper IX - Part 2), which is due to thermal feedback and the superbubble scheme heating up the gas and keeping it from settling into a disc.

This unusual delayed starburst of CHANGA's \textit{target} merger also carries an implication for observational studies. In large surveys studying galaxy mergers, some galaxies may experience little enhancement in SFR during the interaction, even for major mergers. Conversely, the peak merger-driven starburst happens in the post-coalescence stage when the remnant already becomes relaxed, implying that some galaxies may display highly enhanced SFR despite showing no morphological disturbances. This may introduce biases to the control sample in these surveys and statistically weaken the relationship between mergers and SFR enhancement. Hence, observational studies can potentially underestimate the role of mergers in driving starbursts when relying solely on instantaneous structural disturbance indicators, leading to an incomplete interpretation of galaxy mass assembly.




\bibliography{sample631}{}
\bibliographystyle{aasjournal}



\end{document}